\documentclass[11pt]{article}
\usepackage[colorlinks=true,citecolor=blue,linkcolor=blue]{hyperref}
\usepackage{multirow}
\usepackage[left=2cm,right=2cm,top=2cm,bottom=2cm]{geometry}
\usepackage{cite}
\usepackage{psfrag}
\usepackage{graphicx}
\usepackage{graphics}
\usepackage{epsfig}
\usepackage{amsmath,amsfonts,amssymb}
\usepackage{pstcol,pst-fill,pst-grad}
\usepackage{euscript}
\usepackage{pstricks}
\usepackage{wrapfig}
\usepackage{slashed}
\usepackage[toc,page]{appendix}
\usepackage{here}

\date{}
\begin{document}
	\title{\vspace{-3cm}
		\hfill\parbox{4cm}{\normalsize \emph{}}\\
		\vspace{1cm}
		{ Elastic electron-proton scattering in the presence of a circularly polarized laser field}}
	\vspace{2cm}
	
	\author{ I Dahiri$^{1}$, M Jakha$^{1}$, S Mouslih$^{2,1}$,  B Manaut$^{1,}$, S Taj$^1$ \thanks{Corresponding author, E-mail: s.taj@usms.ma}  \\
		{\it {\small$^1$ Sultan Moulay Slimane University, Polydisciplinary Faculty,}}\\
		{\it {\small Research Team in Theoretical Physics and Materials (RTTPM), Beni Mellal, 23000, Morocco.}}\\			
	{\it {\small$^2$Faculty of Sciences and Techniques,
		Laboratory of Materials Physics (LMP),
		Beni Mellal, 23000, Morocco.}}		
	}
	\maketitle \setcounter{page}{1}

\date{\today}
\begin{abstract}
Owing to recent advances in laser technology, it has become important to investigate fundamental laser-assisted processes in very powerful laser fields. In the present work and within the framework of laser-assisted quantum electrodynamics (QED), electron-proton scattering was considered in the presence of a strong electromagnetic field of circular polarization. First, we present a study of the process where we only take into account the relativistic dressing of the electron without the proton. Then, in order to explore the effect of the proton dressing, we fully consider the relativistic dressing of the electron and the proton together and describe them by using Dirac-Volkov functions. The analytical expression for the differential cross section (DCS) in both cases is derived at lowest-order of perturbation theory. As a result, the DCS is notably reduced by the laser field. It is found that the effect of proton dressing begins to appear at laser field strengths greater than or equal to $10^{10}~\text{V/cm}$ and it therefore must be taken into account. The influence of the laser field strength and frequency on the DCS is reported. A comparison with the Mott scattering and the laser-free results is also included.
\end{abstract}
Keywords: QED processes, Laser-assisted, Elastic scattering.

\section{Introduction}
The electromagnetic force is one of the four basic forces that govern our universe, along with gravity, strong and weak nuclear forces, and it is responsible for the attraction or repulsion of charged bodies.  If we can understand how the particles that make up our universe interact with the electromagnetic field and how their properties change during its presence, we will be able to understand very well the radiation-matter interaction and we will also be able to discover new properties of the particles. This will enable us to gather as much information as possible about this universe and exactly how the matter that composes it behaves. Since its invention in the $1960$s, the laser, which is a new light source with excellent monochromaticity, high brightness, strong direction and coherence, has become widely used in various daily applications due to the special properties of its radiation. It is currently considered an indispensable tool for investigating physical processes in various fields covering atomic and plasma physics as well as nuclear and high-energy physics. Charged particle collisions in the presence of laser have attracted a lot of interest in recent decades because of their broad applications and their contribution to the fundamental understanding of the atomic structure in detail. The vast advances in laser technology in recent times have contributed to the emergence of extremely powerful laser sources, with intensities of up to $10^{22}~\text{W/cm}^{2}$ \cite{highlaser}, and as a result, renewed theoretical attention has been given to the study of the basic processes in the presence of a laser field. In particular, electron scattering processes have played a fundamental role in the development of science, both from a theoretical and experimental point of view. Electron scattering is a basic physical process whose importance has long been recognized in atomic and molecular physics. The development of high-power femtosecond lasers has made it possible to experimentally probe these scattering processes and to observe the multiple photon processes. The physics of laser-assisted electron scattering has been extensively reported in the scientific literature. In order to find an overview of this area of research, we refer the reader to several textbooks \cite{faisal,mittleman,fedorov} as well as to some review papers \cite{review2019,review2009,review1990,review1998} and references therein. The majority of these studies focused on the non-relativistic aspects of such processes and at low or moderate field intensities. However, with the use of ultra-powerful lasers, the non-relativistic approach is no longer valid and therefore the relativistic treatment is necessary even in the case of non-relativistic electrons. In some early works on the scattering of electrons by a potential in a strong electromagnetic plane wave \cite{nospin1,nospin2,nospin3}, the effects of the electron spin were neglected, and the electron was considered as a laser-dressed Klein-Gordon particle instead of a Dirac particle. With regard to the relativistic treatment of electron scattering processes, the spin effects of electron are investigated in Mott scattering in the presence of a strong circularly polarized laser field by Szymanowski \textit{et al.} \cite{szymanowski1,szymanowski2}. However, as they also pointed out, the scattering process in this configuration is not so efficient and rich in detail as in the case of linear or elliptical polarization. The same process is treated by Li \textit{et al.} in the presence of a linearly polarized laser field \cite{li2003} as well as in the Coulomb approximation \cite{li2004}. The case of elliptical polarization has been considered by Attaourti \textit{et al.} \cite{attaourti2004}. The effects of the spin polarization in Mott scattering are investigated by Manaut \textit{et al.} \cite{manaut2009,manaut2005}. The resonances in laser-assisted M\o ller scattering were analyzed in the presence of a powerful laser field in \cite{panek2004}. The scattering of an electron by a positron in the light wave field was studied in the works \cite{bhabha1,bhabha2}. The scattering of an electron by a muon has been studied in the first Born approximation in the presence of a laser field with linear \cite{du2018} or elliptical polarization \cite{muon1,muon2,muon3}. Furthermore, several authors have studied other interesting processes, whether in weak interactions such as particle decays \cite{jakha,mouslih,dicus1,dicus2,liu2007} and the elastic scattering of a muon neutrino by an electron in the presence of laser field \cite{neutrino}, or in quantum electrodynamics such as laser-assisted bremsstrahlung\cite{brems1,brems2} and pair production \cite{pairprod1,pairprod2}. The present work is devoted to the electron-proton scattering in the presence of a circularly polarized electromagnetic field. Electron-proton scattering is a basic example that is part of lepton-hadron scattering, which is another type that is no less important than the other types of electron scattering. The study of lepton-hadron scattering at high energies and large momentum transfers allows us to extract more detailed information on the internal structure of hadrons. The electron-proton scattering differs from that of Mott by the fact that the target (proton) is freely movable and therefore the recoil effect must be taken into account. In this paper, we assumed the proton to be a Dirac particle which has no internal structure (point-like particle) and that the energy of the incoming electron was not high enough so that the internal structure of the proton could be detected. At an early stage, Rosenbluth discussed in detail the theory of elastic electron-proton scattering at very high energies (several $100~\text{MeV}$) and found that the formula of the differential cross section (DCS) has to be modified by introducing electric and magnetic form factors representing the internal structure of the proton \cite{rosenbluth}. After that, many calculations were performed to apply the radiative corrections \cite{corrections1,corrections2,corrections3}. To the best of our knowledge, only two articles have addressed the electron-proton scattering, where the proton is not considered to be fixed, in the presence of an electromagnetic field. In $2014$ and in a brief report, the recoil effect in relativistic scattering of an electron from a freely movable proton in the presence of a linearly polarized laser field has been investigated in the first Born approximation by Liu and Li \cite{liu2014}. In another recent paper, Wang \textit{et al.} studied the electron scattering from freely movable proton and positive muon $\mu^{+}$ in the presence of a radiation field and examined the dependencies of the DCS on the laser field intensity and the electron-impact energy \cite{wang2019}. In these two studies, it was found that the laser significantly contributed to the enhancement of the DCS. Apart from this, some early studies dealt with the same subject and considered the proton to be static as an atom or a positive ion H$^{+}$ \cite{ion,atom}. In this paper, we will address the circular polarization case of the laser field and show how could the change of this polarization affect the process during particle scattering. Our main contribution through this is that we have considered the dressing of the proton and explored its effects on the DCS. The remainder of the paper is structured as follows: we begin, in section \ref{sec:absence}, with the most basic results of the electron-proton scattering in the absence of the laser field in order to map the main steps to be followed during our treatment and to avoid distracting the reader by referring to other references for laser-free results. In section \ref{secwith}, we consider the electron-proton process in the presence of circularly polarized monochromatic laser field without dressing of proton, we developed the detailed theoretical formalism required to obtain the DCS. In section \ref{dressing}, in addition to the dressing of the incident and scattered electrons, we take also into account the dressing of the proton which introduces new modifications on the DCS. The results obtained in all cases are presented and discussed in section \ref{sec:res}. Finally, our conclusions are summarized in section \ref{sec:conclusion}. We note here that we have used, throughout this work, the metric tensor $g^{\mu\nu}=\text{diag}(1,-1,-1,-1)$ and atomic units in which one has $\hbar=m_{e}=e=1$ where $m_{e}$ is the rest mass of the electron. For all $k$, the bold style $\textbf{k}$ is reserved for vectors.
\section{Electron-proton scattering in the absence of a laser field}\label{sec:absence}
This $e$-$P$ scattering can be expressed as
\begin{equation}\label{process}
e^{-}(p_{1})+P(p_{2})\longrightarrow e^{-}(p_{3})+P(p_{4}),
\end{equation}
the labels $p_{i}$ are the associated momenta, where the odd indexes $(i=1,3)$ stand for electron while the even ones $(i=2,4)$ stand for proton. We follow the usual steps of calculations and give the S-matrix element corresponding to this process in lowest order:
\begin{equation}\label{smatrix0}
S_{fi}=-i\int d^{4}x~\overline{\psi}_{p_{3}}(x)\slashed{A}(x)\psi_{p_{1}}(x),
\end{equation}
where $\psi_{p_{1}}(x)$ and $\psi_{p_{3}}(x)$ are the plane Dirac waves that describe, respectively, the initial and final electrons and can be written, when normalized to the volume $V$, as follows:
\begin{equation}\label{ewaves}
\begin{split}
\psi_{p_{1}}(x)&=\frac{1}{\sqrt{2E_{1}V}}u(p_{1},s_{1})e^{-ip_{1}.x},\\
\psi_{p_{3}}(x)&=\frac{1}{\sqrt{2E_{3}V}}u(p_{3},s_{3})e^{-ip_{3}.x}.
\end{split}
\end{equation}
$\slashed{A}(x)=A_{\mu}(x)\gamma^{\mu}$ with $A_{\mu}(x)$ is the four-potential produced by the proton and it has the form
\begin{equation}
A_{\mu}(x)=-\int d^{4}y~D_{F}(x-y)\overline{\psi}_{p_{4}}(y)\gamma_{\mu}\psi_{p_{2}}(y),
\end{equation}
where $D_{F}(x-y)$ is the Feynman propagator for the electromagnetic radiation given by \cite{greiner}:
\begin{equation}
D_{F}(x-y)=\int\frac{d^{4}q}{(2\pi)^{4}}e^{-iq(x-y)}\bigg(\frac{-1}{q^{2}+i\varepsilon}\bigg).
\end{equation}
Then
\begin{equation}\label{potential}
A_{\mu}(x)=\int d^{4}y~\frac{d^{4}q}{(2\pi)^{4}}\frac{e^{-iq(x-y)}}{q^{2}+i\varepsilon}\big[\overline{\psi}_{p_{4}}(y)\gamma_{\mu}\psi_{p_{2}}(y)\big],
\end{equation}
where $\psi_{p_{2}}(y)$ and $\psi_{p_{4}}(y)$ are the free initial and final states of the proton and they have the same form as the electron waves (\ref{ewaves}):
\begin{equation}\label{pwaves}
\begin{split}
\psi_{p_{2}}(y)&=\frac{1}{\sqrt{2E_{2}V}}u(p_{2},s_{2})e^{-ip_{2}.y},\\
\psi_{p_{4}}(y)&=\frac{1}{\sqrt{2E_{4}V}}u(p_{4},s_{4})e^{-ip_{4}.y}.
\end{split}
\end{equation}
In equations (\ref{ewaves}) and (\ref{pwaves}), $s_{i}$ and  $E_{i}$ $(i=1,2,3,4)$ denote the spin and energy, respectively. Substituting equation (\ref{potential}) and the wave functions into the S-matrix element (\ref{smatrix0}), this reads
\begin{equation}\label{smatrix1}
S_{fi}=\frac{-i}{\sqrt{16E_{1}E_{2}E_{3}E_{4}V^{4}}}\int d^{4}x~d^{4}y~\frac{d^{4}q}{(2\pi)^{4}}\frac{e^{i(p_{3}-p_{1}-q).x}e^{i(p_{4}-p_{2}+q).y}}{q^{2}+i\varepsilon}\mathcal{M}_{fi},
\end{equation}
where
\begin{equation}
\mathcal{M}_{fi}=\big[\overline{u}(p_{3},s_{3})\gamma^{\mu}u({p_{1},s_{1}})\big]\big[\overline{u}(p_{4},s_{4})\gamma_{\mu}u({p_{2},s_{2}})\big].
\end{equation}
The integration over four-dimensional spatial coordinates $d^{4}x$ and $d^{4}y$ can be carried out at once, resulting in
\begin{equation}
\begin{split}
\int d^{4}x~e^{i(p_{3}-p_{1}-q).x}&=(2\pi)^{4}\delta^{4}(p_{3}-p_{1}-q),\\
\int d^{4}y~e^{i(p_{4}-p_{2}+q).y}&=(2\pi)^{4}\delta^{4}(p_{4}-p_{2}+q),
\end{split}
\end{equation}
and the reminder integration over $d^{4}q$ can be performed simply as follows:
\begin{equation}
\int \frac{d^{4}q}{(2\pi)^{4}}\frac{(2\pi)^{4}\delta^{4}(p_{3}-p_{1}-q)(2\pi)^{4}\delta^{4}(p_{4}-p_{2}+q)}{q^{2}+i\varepsilon}=\frac{(2\pi)^{4}\delta^{4}(p_{4}-p_{2}+p_{3}-p_{1})}{(p_{3}-p_{1})^{2}+i\varepsilon},
\end{equation}
and the total S-matrix element (\ref{smatrix1}) reads
\begin{equation}
S_{fi}=\frac{-i}{\sqrt{16E_{1}E_{2}E_{3}E_{4}V^{4}}}\frac{(2\pi)^{4}\delta^{4}(p_{4}-p_{2}+p_{3}-p_{1})}{q^{2}+i\varepsilon}\mathcal{M}_{fi},
\end{equation}
where $q=p_{3}-p_{1}$ is the relativistic four-momentum transfer in the absence of the laser field. To calculate
the scattering cross-section, we multiply the squared S-matrix
element $|S_{fi}|^{2}$  by the density of final states, and divide by the observation time interval $T$ and the flux of the incoming particles $|J_{\text{inc.}}|$ and finally one has to average over the initial spins and to sum over the final ones. We obtain
\begin{equation}
\begin{split}
d\overline{\sigma}&=\frac{Vd^{3}p_{3}}{(2\pi)^{3}}\frac{Vd^{3}p_{4}}{(2\pi)^{3}}\frac{1}{|J_{\text{inc.}}|T}\frac{(2\pi)^{4}VT\delta^{4}(p_{4}-p_{2}+p_{3}-p_{1})}{16E_{1}E_{2}E_{3}E_{4}V^{4}q^{4}}|\overline{\mathcal{M}_{fi}}|^{2},
\end{split}
\end{equation}
where
\begin{equation}
|\overline{\mathcal{M}_{fi}}|^{2}=\frac{1}{4}\sum_{s_{i}}|\mathcal{M}_{fi}|^{2}.
\end{equation}
With the help of the following formula \cite{greiner}
\begin{equation}\label{formula0}
\frac{d^{3}p_{4}}{E_{4}}=2\int_{-\infty}^{+\infty}d^{4}p_{4}\delta(p_{4}^{2}-M^{2}c^{2})\Theta(p_{4}^{0}),
\end{equation}
with
\[\Theta(p_{4}^{0})=\left\lbrace\begin{array}{ccc}
&1&~~\text{for}~~p_{4}^{0}>0\\
&0&~~\text{for}~~p_{4}^{0}<0
\end{array}\right.\]
and $M=1836.152~672~45 \times m_{e}$ is the rest mass of the proton, the cross-section becomes
\begin{equation}
\begin{split}
d\overline{\sigma}=\frac{1}{8(2\pi)^{2}E_{1}E_{2}E_{3}|J_{\text{inc.}}|Vq^{4}}\int \delta(p_{4}^{2}-M^{2}c^{2})d^{3}p_{3}|\overline{\mathcal{M}_{fi}}|^{2},
\end{split}
\end{equation}
with $p_{4}+p_{3}-p_{1}-p_{2}=0$. By considering the flux incident of electron in the rest frame of proton $|J_{\text{inc.}}|=|\textbf{p}_{1}|c^{2}/E_{1}V$ and using $d^{3}p_{3}=|\textbf{p}_{3}|^{2}d|\textbf{p}_{3}|d\Omega_{f}$ and $E_{3}dE_{3}=c^{2}|\textbf{p}_{3}|d|\textbf{p}_{3}|$, we get
\begin{equation}
\frac{d\overline{\sigma}}{d\Omega_{f}}=\frac{1}{8(2\pi)^{2}c^{6}Mq^{4}}\frac{|\textbf{p}_{3}|}{|\textbf{p}_{1}|}\int \delta(p_{4}^{2}-M^{2}c^{2})dE_{3}|\overline{\mathcal{M}_{fi}}|^{2}.
\end{equation}
The integration over $dE_{3}$ can be performed using the following familiar formula \cite{greiner}
\begin{align}\label{familiarformula}
\int dxf(x)\delta(g(x))=\dfrac{f(x)}{|g'(x)|}\bigg|_{g(x)=0}.
\end{align}
Thus we get
\begin{equation}\label{dcswithout}
\bigg(\frac{d\overline{\sigma}}{d\Omega_{f}}\bigg)^{\text{laser-free}}=\frac{1}{16(2\pi)^{2}Mc^{6}q^{4}}\frac{|\textbf{p}_{3}|}{|\textbf{p}_{1}|}\frac{|\overline{\mathcal{M}_{fi}}|^{2}}{|g'(E_{3})|},
\end{equation}
where
\begin{equation}
g'(E_{3})=M+\frac{E_{1}}{c^{2}}-\frac{E_{3}|\textbf{p}_{1}|}{c^{2}|\textbf{p}_{3}|}F(\theta_{i},\theta_{f},\varphi_{i},\varphi_{f}),
\end{equation}
with
\begin{equation}
\begin{split}
F(\theta_{i},\theta_{f},\varphi_{i},\varphi_{f})=&\cos(\varphi_{i})\sin(\theta_{i})\cos(\varphi_{f})\sin(\theta_{f})+\sin(\theta_{i})\sin(\varphi_{i})\sin(\theta_{f})\sin(\varphi_{f})\\&+\cos(\theta_{i})\cos(\theta_{f}),
\end{split}
\end{equation}
and
\begin{equation}
|\overline{\mathcal{M}_{fi}}|^{2}=\frac{1}{4}Tr[(c\slashed{p}_{3}+c^{2})\gamma^{\mu}(c\slashed{p}_{1}+c^{2})\gamma^{\nu}]Tr[(c\slashed{p}_{4}+Mc^{2})\gamma_{\mu}(c\slashed{p}_{2}+Mc^{2})\gamma_{\nu}].
\end{equation}
\section{Electron-proton scattering in the presence of a laser field with circular polarization}\label{secwith}
Let us now consider the $e$-$P$ process (\ref{process}) and assume that this scattering occurs in the presence of a circularly polarized monochromatic laser field, which is described by its classical four-potential $A(\phi)$ that satisfies the Lorentz gauge condition $\partial A(\phi)=0$. It is expressed as follows:
\begin{equation}\label{potlaser}
A_{\text{laser}}^{\mu}(\phi)=a^{\mu}_{1}\cos(\phi)+a^{\mu}_{2}\sin(\phi),~~~\phi=(k.x),
\end{equation}
where $k=(\omega/c,\textbf{k})$ is the wave 4-vector $(k^{2}=0)$, $\phi$ is the phase of the laser field and $\omega$ its frequency. The polarization 4-vectors $a^{\mu}_{1}$ and $a^{\mu}_{2}$ are equal in magnitude and orthogonal such that:
\begin{equation}
\begin{split}
a^{\mu}_{1}&=(0,|\text{\textbf{a}}|,0,0),\\
a^{\mu}_{2}&=(0,0,|\text{\textbf{a}}|,0),
\end{split}
\end{equation}
which implies $(a_{1}.a_{2})=0$ and $a_{1}^{2}=a_{2}^{2}=a^{2}=-|\text{\textbf{a}}|^{2}=-(c\mathcal{E}_{0}/\omega)^{2}$ where $\mathcal{E}_{0}$ is the laser field strength and $c\simeq137~\text{a.u}$ is the velocity of light in the vacuum. The Lorentz gauge condition applied to the four-potential yields
\begin{equation}
k_{\mu}A^{\mu}=0,
\end{equation}
which implies $(k.a_{1})=(k.a_{2})=0$, meaning that the wave vector $\textbf{k}$ is chosen to be along the $z$-axis. In such conditions, the electron state in the laser field is described by the following relativistic Dirac-Volkov functions \cite{volkov}, where $i=1,3$:
\begin{equation}
\psi_{p_{i}}(x)=\bigg(1+\frac{\slashed{k}\slashed{A}}{2c(k.p_{i})}\bigg)\frac{u(p_{i},s_{i})}{\sqrt{2Q_{i}V}}e^{iS(q_{i},x)},
\end{equation}
where
\begin{equation}
S(q_{i},x)=-q_{i}.x-\frac{a_{1}.p_{i}}{c(k.p_{i})}\sin(\phi)+\frac{a_{2}.p_{i}}{c(k.p_{i})}\cos(\phi).
\end{equation}
$u(p_{i},s_{i})$ is the free electron bispinor normalized as $\bar{u}(p_{i},s_{i})u(p_{i},s_{i})=2c^{2}$, and $q_{i}=(Q_{i}/c,\textbf{q}_{i})$ is the dressed four-momentum acquired by the electron in the presence of the laser field
\begin{equation}
q_{i}=p_{i}-\frac{a^{2}}{2c^{2}(k.p_{i})}k.
\end{equation}
The square of this momentum is
\begin{equation}
q_{i}^{2}=m_{*}^{2}c^{2},
\end{equation}
where $m_{*}$ is an effective mass of the electron inside the electromagnetic field
\begin{equation}\label{meffe}
m_{*}=\sqrt{1-\frac{a^{2}}{c^{4}}}.
\end{equation}
The proton, due to its large mass, is much less influenced by the laser than the electron (at least, for the laser field strengths considered here). It will be treated, in the first Born approximation, as a structureless Dirac (spin-$\tfrac{1}{2}$) particle. Therefore, we will describe it in the first stage using the same previous unaffected plane waves given in equation~(\ref{pwaves}). Then, the S-matrix element (\ref{smatrix0}) becomes
\begin{equation}\label{smatrixlaser}
\begin{split}
S_{fi}=&\frac{-i}{\sqrt{16Q_{1}Q_{3}E_{2}E_{4}V^{4}}}\int d^{4}x~d^{4}y~\frac{d^{4}q}{(2\pi)^{4}}\frac{e^{i(S(q_{1},x)-S(q_{3},x)-q.x)}e^{i(p_{4}-p_{2}+q).y}}{q^{2}+i\varepsilon}\\
&\times \bigg[\overline{u}(p_{3},s_{3})\bigg(1+\frac{\slashed{A}\slashed{k}}{2c(k.p_{3})}\bigg)\gamma^{\mu}\bigg(1+\frac{\slashed{k}\slashed{A}}{2c(k.p_{1})}\bigg)u(p_{1},s_{1})\bigg]\bigg[\overline{u}(p_{4},s_{4})\gamma_{\mu}u(p_{2},s_{2})\bigg].
\end{split}
\end{equation}
The exponential term $e^{i(S(q_{1},x)-S(q_{3},x))}$ can be recasted in a suitable form by introducing the following parameters
\begin{equation}\label{argument}
z=\sqrt{\alpha_{1}^{2}+\alpha_{2}^{2}}~~~\text{and}~~~\phi_{0}=\arctan(\alpha_{2}/ \alpha_{1}),
\end{equation}
where
\begin{equation}
\begin{split}
\alpha_{1}=&\frac{a_{1}.p_{1}}{c(k.p_{1})}-\frac{a_{1}.p_{3}}{c(k.p_{3})},\\
\alpha_{2}=&\frac{a_{2}.p_{1}}{c(k.p_{1})}-\frac{a_{2}.p_{3}}{c(k.p_{3})},
\end{split}
\end{equation}
then we can write
\begin{equation}
e^{i(S(q_{1},x)-S(q_{3},x))}=e^{i(q_{3}-q_{1}).x}e^{-iz\sin(\phi-\phi_{0})}.
\end{equation}
Therefore, the S-matrix element becomes
\begin{equation}\label{smatrix2}
\begin{split}
S_{fi}=&\frac{-i}{\sqrt{16Q_{1}Q_{3}E_{2}E_{4}V^{4}}}\int d^{4}x~d^{4}y~\frac{d^{4}q}{(2\pi)^{4}}\frac{e^{i(q_{3}-q_{1}-q).x}e^{i(p_{4}-p_{2}+q).y}e^{-iz\sin(\phi-\phi_{0})}}{q^{2}+i\varepsilon}\\
&\times \big[\overline{u}(p_{3},s_{3})\big(C^{0}_\mu+C^{1}_\mu\cos(\phi)+C^{2}_\mu\sin(\phi)\big)u(p_{1},s_{1})\big]\big[\overline{u}(p_{4},s_{4})\gamma^{\mu}u(p_{2},s_{2})\big],
\end{split}
\end{equation}
where
\begin{equation}
\begin{split}
C^{0}_\mu&=\gamma_{\mu}-2k_{\mu}a^{2}\slashed{k}C(p_{1})C(p_{3}),\\
C^{1}_\mu&=C(p_{1})\gamma_{\mu}\slashed{k}\slashed{a}_{1}+C(p_{3})\slashed{a}_{1}\slashed{k}\gamma_{\mu},\\
C^{2}_\mu&=C(p_{1})\gamma_{\mu}\slashed{k}\slashed{a}_{2}+C(p_{3})\slashed{a}_{2}\slashed{k}\gamma_{\mu},
\end{split}
\end{equation}
with $C(p_{i})=1/2c(k.p_{i})$. The three  different  quantities  in equation~(\ref{smatrix2}) can be transformed by the well-known identities involving ordinary Bessel functions $J_{s}(z)$, where $z$ is their argument defined in equation~(\ref{argument}) and $s$, their ordre, will be interpreted in the following as the number of exchanged photons. Using such transformations \cite{landau} in equation~(\ref{smatrix2}), the matrix element $S_{fi}$ becomes
\begin{equation}\label{smatrix3}
\begin{split}
S_{fi}=&\frac{-i}{\sqrt{16Q_{1}Q_{3}E_{2}E_{4}V^{4}}}\sum_{s=-\infty}^{+\infty}\int d^{4}x~d^{4}y~\frac{d^{4}q}{(2\pi)^{4}}\frac{e^{i(q_{3}-q_{1}-sk-q).x}e^{i(p_{4}-p_{2}+q).y}}{q^{2}+i\varepsilon} \\
&\times \big[\overline{u}(p_{3},s_{3})\big(C^{0}_\mu B_{s}(z)+C^{1}_\mu B_{1s}(z)+C^{2}_\mu B_{2s}(z)\big)u(p_{1},s_{1})\big]\big[\overline{u}(p_{4},s_{4})\gamma^{\mu}u(p_{2},s_{2})\big].
\end{split}
\end{equation}
where
\begin{align}
\begin{split}
\begin{bmatrix}
B_{s}(z)\\
B_{1s}(z)\\
B_{2s}(z) \end{bmatrix}=\begin{bmatrix}J_{s}(z)e^{is\phi_{0}}\\
\big(J_{s+1}(z)e^{i(s+1)\phi_{0}}+J_{s-1}(z)e^{i(s-1)\phi_{0}}\big)/2\\
\big(J_{s+1}(z)e^{i(s+1)\phi_{0}}-J_{s-1}(z)e^{i(s-1)\phi_{0}}\big)/2i
 \end{bmatrix}.
\end{split}
\end{align}
Integrating over $d^{4}x$, $d^{4}y$ and $d^{4}q$, we get
\begin{equation}\label{smatrix4}
\begin{split}
S_{fi}=&\frac{-i}{\sqrt{16Q_{1}Q_{3}E_{2}E_{4}V^{4}}}\sum_{s=-\infty}^{+\infty}\frac{(2\pi)^{4}\delta(p_{4}-p_{2}+q_{3}-q_{1}-sk)}{q^{2}+i\varepsilon}\\
&\times \big[\overline{u}(p_{3},s_{3})\big(C^{0}_\mu B_{s}(z)+C^{1}_\mu B_{1s}(z)+C^{2}_\mu B_{2s}(z)\big)u(p_{1},s_{1})\big]\big[\overline{u}(p_{4},s_{4})\gamma^{\mu}u(p_{2},s_{2})\big],
\end{split}
\end{equation}
with $q=q_{3}-q_{1}-sk$ is the relativistic four-momentum transfer in the presence of the laser field. Because electron-proton scattering experiments are often carried out with a fixed target in the laboratory frame, we will evaluate the differential cross section (DCS) in the rest frame of the proton by following the same steps as in the absence of the laser field in the previous section. This yields
\begin{equation}
\begin{split}
d\overline{\sigma}&=\frac{Vd^{3}q_{3}}{(2\pi)^{3}}\frac{Vd^{3}p_{4}}{(2\pi)^{3}}\frac{1}{|J_{\text{inc.}}|T}\frac{(2\pi)^{4}VT\delta^{4}(p_{4}-p_{2}+q_{3}-q_{1}-sk)}{16Q_{1}Q_{3}E_{2}E_{4}V^{4}q^{4}}|\overline{\mathcal{M}_{fi}^{s}}|^{2}.
\end{split}
\end{equation}
With the use of the formula (\ref{formula0}) and the flux $|J_{\text{inc.}}|=|\textbf{q}_{1}|c^{2}/Q_{1}V$, we get
\begin{equation}\label{dcswith}
\bigg(\frac{d\overline{\sigma}}{d\Omega_{f}}\bigg)^{\text{with laser}}=\sum_{s=-\infty}^{+\infty}\frac{d\overline{\sigma}^{s}}{d\Omega_{f}}=\frac{1}{16(2\pi)^{2}Mc^{6}q^{4}}\sum_{s=-\infty}^{+\infty}\frac{|\textbf{q}_{3}|}{|\textbf{q}_{1}|}\frac{|\overline{\mathcal{M}^{s}_{fi}}|^{2}}{|g'(Q_{3})|},
\end{equation}
with $p_{4}-p_{2}+q_{3}-q_{1}-sk=0$, and where
\begin{equation}
g'(Q_{3})=\bigg[\frac{Q_{1}}{c^{2}}+M+\frac{s\omega}{c^{2}}\bigg]-\frac{Q_{3}}{c^{2}|\textbf{q}_{3}|}\bigg[|\textbf{q}_{1}|F(\theta_{i},\theta_{f},\varphi_{i},\varphi_{f})+\frac{s\omega}{c}\cos(\theta_{f})\bigg],
\end{equation}
and
\begin{equation}\label{trace}
|\overline{\mathcal{M}^{s}_{fi}}|^{2}=\frac{1}{4}Tr[(c\slashed{p}_{3}+c^{2})\Gamma^{s}_{\mu}(c\slashed{p}_{1}+c^{2})\overline{\Gamma}^{s}_{\nu}]Tr[(c\slashed{p}_{4}+Mc^{2})\gamma^{\mu}(c\slashed{p}_{2}+Mc^{2})\gamma^{\nu}],
\end{equation}
where
\begin{equation}
\Gamma^{s}_{\mu}=C^{0}_{\mu}B_{s}(z)+C^{1}_{\mu}B_{1s}(z)+C^{2}_{\mu}B_{2s}(z),
\end{equation}
and
\begin{align}
\begin{split}
\bar{\Gamma}^{s}_{\nu}&=\gamma^{0}\Gamma_{\nu}^{s\dagger}\gamma^{0},\\
&=\overline{C}^{0}_{\nu}B^{*}_{s}(z)+ \overline{C}^{1}_{\nu}B^{*}_{1s}(z)+\overline{C}^{2}_{\nu}B^{*}_{2s}(z),
\end{split}
\end{align}
where
\begin{equation}
\begin{split}
\overline{C}^{0}_{\nu}=&\gamma^{0}C^{0\dagger}_{\nu}\gamma^{0}=\gamma_{\nu}-2k_{\nu}a^{2}\slashed{k}C(p_{1})C(p_{3}),\\
\overline{C}^{1}_{\nu}=&\gamma^{0}C^{1\dagger}_{\nu}\gamma^{0}=C(p_{1})\slashed{a}_{1}\slashed{k}\gamma_{\nu}+C(p_{3})\gamma_{\nu}\slashed{k}\slashed{a}_{1},\\
\overline{C}^{2}_{\nu}=&\gamma^{0}C^{2\dagger}_{\nu}\gamma^{0}=C(p_{1})\slashed{a}_{2}\slashed{k}\gamma_{\nu}+C(p_{3})\gamma_{\nu}\slashed{k}\slashed{a}_{2}.
\end{split}
\end{equation}
By using the FeynCalc Mathematica package designed for the trace calculations \cite{feyncalc1,feyncalc2,feyncalc3}, we can comfortably program and calculate numerically the two traces appearing above in equation (\ref{trace}). The result we have obtained is as follows
\begin{equation}\label{spinorpart}
\begin{split}
|\overline{\mathcal{M}^{s}_{fi}}|^{2}=&M_{1}|B_{s}|^{2}+M_{2}|B_{1s}|^{2}+M_{3}|B_{2s}|^{2}+M_{4}B_{s}B^{*}_{1s}+M_{5}B_{s}B^{*}_{2s}\\
&+M_{6}B_{1s}B^{*}_{s}+M_{7}B_{1s}B^{*}_{2s}+M_{8}B_{2s}B^{*}_{s}+M_{9}B_{2s}B^{*}_{1s},
\end{split}
\end{equation}
where the argument $z$ of the various ordinary Bessel functions has been dropped for convenience and the nine coefficients $M_{1}$, $M_{2}$, $M_{3}$, $M_{4}$, $M_{5}$, $M_{6}$, $M_{7}$, $M_{8}$ and $M_{9}$ are explicitly expressed, in terms of different scalar products, by
\begin{equation}
\begin{split}
M_{1}=&\dfrac{4}{(k.p_{1}) (k.p_{3})}\big[a^4 (k.p_{2}) (k.p_{4}) - a^2 c^2 (2 c^2 ((k.p_{2}) (k.p_{4}) + (k.p_{1}) (k.p_{3}) M^2)+(k.p_{3})\\
&\times(k.p_{4}) (p_{1}.p_{2}) - 2 (k.p_{2}) (k.p_{4}) (p_{1}.p_{3})+(k.p_{2}) (k.p_{3}) (p_{1}.p_{4}) + (k.p_{1}) (k.p_{4}) (p_{2}.p_{3})\\
&-2 (k.p_{1}) (k.p_{3}) (p_{2}.p_{4}) + (k.p_{1}) (k.p_{2}) (p_{3}.p_{4})) +2 c^4 (k.p_{1}) (k.p_{3}) (2 c^4 M^2 + (p_{1}.p_{4})\\
&\times(p_{2}.p_{3}) - c^2 (M^2 (p_{1}.p_{3}) + (p_{2}.p_{4})) + (p_{1}.p_{2}) (p_{3}.p_{4}))\big],
\end{split}
\end{equation}
\begin{equation}
\begin{split}
M_{2}=&\dfrac{4 c^2 }{(k.p_{1}) (k.p_{3})}\big[ -2 (a_{1}.p_{1}) (a_{1}.p_{4}) (k.p_{2}) (k.p_{3}) + (a_{1}.p_{3}) (-2 (a_{1}.p_{4}) (k.p_{1}) (k.p_{2})\\
& + 4 (a_{1}.p_{1}) (k.p_{2}) (k.p_{4})) + a^2 (c^2 (2 (k.p_{2}) (k.p_{4}) + ((k.p_{1})^2 + (k.p_{3})^2) M^2) + (k.p_{3})\\
&\times (k.p_{4}) (p_{1}.p_{2}) - 2 (k.p_{2}) (k.p_{4}) (p_{1}.p_{3}) + (k.p_{2}) (k.p_{3}) (p_{1}.p_{4}) - (k.p_{3}) (k.p_{4}) (p_{2}.p_{3}) \\
& - (k.p_{2}) (k.p_{3}) (p_{3}.p_{4})+ (k.p_{1}) (-(k.p_{4}) (p_{1}.p_{2}) - (k.p_{2}) (p_{1}.p_{4}) + (k.p_{4}) (p_{2}.p_{3}) \\
&- 2 (k.p_{3}) (p_{2}.p_{4}) + (k.p_{2}) (p_{3}.p_{4})))\big],
\end{split}
\end{equation}
\begin{equation}
\begin{split}
M_{3}=&\dfrac{4 c^2 }{(k.p_{1}) (k.p_{3})}\big[ -2 (a_{2}.p_{1}) (a_{2}.p_{4}) (k.p_{2}) (k.p_{3}) + (a_{2}.p_{3}) (-2 (a_{2}.p_{4}) (k.p_{1}) (k.p_{2}) \\
&+ 4 (a_{2}.p_{1}) (k.p_{2}) (k.p_{4})) + a^2 (c^2 (2 (k.p_{2}) (k.p_{4}) + ((k.p_{1})^2 + (k.p_{3})^2) M^2) + (k.p_{3}) \\
&\times(k.p_{4}) (p_{1}.p_{2}) - 2 (k.p_{2}) (k.p_{4}) (p_{1}.p_{3}) + (k.p_{2}) (k.p_{3}) (p_{1}.p_{4}) - (k.p_{3}) (k.p_{4}) (p_{2}.p_{3}) \\
&- (k.p_{2}) (k.p_{3}) (p_{3}.p_{4}) + (k.p_{1}) (-(k.p_{4}) (p_{1}.p_{2}) - (k.p_{2}) (p_{1}.p_{4}) + (k.p_{4}) (p_{2}.p_{3}) \\
&- 2 (k.p_{3}) (p_{2}.p_{4}) + (k.p_{2}) (p_{3}.p_{4})))\big],
\end{split}
\end{equation}
\begin{equation}
\begin{split}
M_{4}=M_{6}=&\dfrac{2 c }{(k.p_{1}) (k.p_{3})}\big[a^2 (k.p_{2}) ((a_{1}.p_{4}) ((k.p_{1}) + (k.p_{3})) - 2 ((a_{1}.p_{1}) + (a_{1}.p_{3})) (k.p_{4}))\\
& + 2 c^2 ((a_{1}.p_{3})(k.p_{1}) (c^2 (-(k.p_{1}) + (k.p_{3})) M^2 + (k.p_{4}) (p_{1}.p_{2}) + (k.p_{2}) (p_{1}.p_{4})) \\
&+ (k.p_{3}) (-(a_{1}.p_{4}) (k.p_{1}) ((p_{1}.p_{2}) + (p_{2}.p_{3}))+(a_{1}.p_{1}) (c^2 ((k.p_{1}) - (k.p_{3})) M^2\\
& + (k.p_{4}) (p_{2}.p_{3}) + (k.p_{2}) (p_{3}.p_{4}))))\big],
\end{split}
\end{equation}
\begin{equation}
\begin{split}
M_{5}=M_{8}=&\dfrac{2 c }{(k.p_{1}) (k.p_{3})}\big[ a^2 (k.p_{2}) ((a_{2}.p_{4}) ((k.p_{1}) + (k.p_{3})) - 2 ((a_{2}.p_{1}) + (a_{2}.p_{3})) (k.p_{4}))\\
&+2 c^2 ((a_{2}.p_{3})(k.p_{1}) (c^2 (-(k.p_{1}) + (k.p_{3})) M^2 + (k.p_{4}) (p_{1}.p_{2}) + (k.p_{2}) (p_{1}.p_{4}))\\
&+ (k.p_{3}) (-(a_{2}.p_{4}) (k.p_{1}) ((p_{1}.p_{2}) + (p_{2}.p_{3}))+ (a_{2}.p_{1}) (c^2 ((k.p_{1}) - (k.p_{3})) M^2\\
& + (k.p_{4}) (p_{2}.p_{3}) + (k.p_{2}) (p_{3}.p_{4}))))\big],
\end{split}
\end{equation}
\begin{equation}
\begin{split}
M_{7}=M_{9}=&\dfrac{-1 }{(k.p_{1}) (k.p_{3})}\big[4 c^2 (k.p_{2}) ((a_{1}.p_{4}) (a_{2}.p_{3}) (k.p_{1}) + (a_{1}.p_{3}) (a_{2}.p_{4}) (k.p_{1}) + (a_{1}.p_{4})\\
&\times (a_{2}.p_{1}) (k.p_{3})+(a_{1}.p_{1}) (a_{2}.p_{4})(k.p_{3})- 2 ((a_{1}.p_{3}) (a_{2}.p_{1}) + (a_{1}.p_{1})\\&\times (a_{2}.p_{3})) (k.p_{4})) \big].
\end{split}
\end{equation}
\section{Proton-dressing effect}\label{dressing}
So far, we only take into account the dressing of the incident and scattered electrons. In this section and in order to realize the laser-dressing effect on the proton, we will take into account the relativistic dressing of both particles involved in this process, the electron and the proton, and thus it will be described together by the Dirac-Volkov plane waves. This will give rise to a new trace to be computed, but it will become clear that taking into account the relativistic dressing of the proton will simply introduce a new sum on the number of photons $l$ that will be exchanged between the proton and the laser field. In this case, the four-potential $A_{\mu}(x)$ produced by the proton becomes
\begin{equation}\label{potential1}
A_{\mu}(x)=\int d^{4}y~\frac{d^{4}q}{(2\pi)^{4}}\frac{e^{-iq(x-y)}}{q^{2}+i\varepsilon}\big[\overline{\psi}_{q_{4}}(y)\gamma_{\mu}\psi_{q_{2}}(y)\big],
\end{equation}
where $\psi_{q_{2}}(y)$ and $\psi_{q_{4}}(y)$ are the Dirac-Volkov plane waves of the initial and final states of the dressed proton. They can be written, where $i=2,4$, as follows:
\begin{equation}\label{pwaveslaser}
\begin{split}
\psi_{p_{i}}(y)&=\bigg(1+\frac{e_{P}\slashed{k}\slashed{A}'}{2c(k.p_{i})}\bigg)\frac{u(p_{i},s_{i})}{\sqrt{2Q_{i}V}}e^{iS(q_{i},y)},\\
\end{split}
\end{equation}
where $e_{P}=-e>0$ is the proton's charge, and $A'_{\text{laser}}(\phi')$ is the four potential of the
laser field felt by the proton
\begin{equation}\label{potlaser}
A_{\text{laser}}^{'\mu}(\phi')=a^{\mu}_{1}\cos(\phi')+a^{\mu}_{2}\sin(\phi'),
\end{equation}
where $\phi'=(k.y)$ is the phase of the laser field. $Q_{i}$ is the total energy acquired by the proton in the presence of a laser field, such that
\begin{equation}
Q_{i}=E_{i}-\frac{a^{2}\omega}{2c^{2}(k.p_{i})}.
\end{equation}
The quantity $S(q_{i},y)$ is given by
\begin{equation}
S(q_{i},y)=-q_{i}.y+\frac{a_{1}.p_{i}}{c(k.p_{i})}\sin(\phi')-\frac{a_{2}.p_{i}}{c(k.p_{i})}\cos(\phi').
\end{equation}
We do not have to repeat the steps leading to the DCS since they have already been presented in the previous sections. Proceeding in the same way as before, we obtain for the unpolarized DCS
\begin{equation}\label{dressedprotondcs}
\bigg(\frac{d\overline{\sigma}}{d\Omega_{f}}\bigg)^{\text{dressed proton}}=\sum_{s,l=-\infty}^{+\infty}\frac{d\overline{\sigma}^{(s,l)}}{d\Omega_{f}}=\frac{1}{16(2\pi)^{2}M^{*}c^{6}q^{4}}\sum_{s,l=-\infty}^{+\infty}\frac{|\textbf{q}_{3}|}{|\textbf{q}_{1}|}\frac{|\overline{\mathcal{M}^{(s,l)}_{fi}}|^{2}}{|h'(Q_{3})|},
\end{equation}
where $M^{*}$ is the effective mass of the proton acquired inside the laser field
\begin{equation}\label{meffp}
M_{*}=\sqrt{M^{2}-\frac{a^{2}}{c^{4}}},
\end{equation}
and $q=q_{3}-q_{1}-sk$. In this case, the energy-momentum conservation $q_{4}+q_{3}-q_{1}-q_{2}-(s+l)k=0$ must be satisfied. The term $|\overline{\mathcal{M}^{(s,l)}_{fi}}|^{2}$ is expressed by
\begin{equation}
|\overline{\mathcal{M}^{(s,l)}_{fi}}|^{2}=\frac{1}{4}Tr[(c\slashed{p}_{3}+c^{2})\Gamma^{s}_{\mu}(c\slashed{p}_{1}+c^{2})\overline{\Gamma}^{s}_{\nu}]Tr[(c\slashed{p}_{4}+Mc^{2})\Lambda_{l}^{\mu}(c\slashed{p}_{2}+Mc^{2})\overline{\Lambda}_{l}^{\nu}],
\end{equation}
with 
\begin{equation}
\Lambda_{l}^{\mu}=C^{'\mu}_{0} B_{l}(z_{P})+C^{'\mu}_{1} B_{1l}(z_{P})+C^{'\mu}_{2} B_{2l}(z_{P}),
\end{equation}
and
\begin{equation}
\begin{split}
C^{'\mu}_{0} &=\gamma^{\mu}-2k^{\mu}a^{2}\slashed{k}C(p_{2})C(p_{4}),\\
C^{'\mu}_{1} &=-C(p_{2})\gamma^{\mu}\slashed{k}\slashed{a}_{1}-C(p_{4})\slashed{a}_{1}\slashed{k}\gamma^{\mu},\\
C^{'\mu}_{2} &=-C(p_{2})\gamma^{\mu}\slashed{k}\slashed{a}_{2}-C(p_{4})\slashed{a}_{2}\slashed{k}\gamma^{\mu}.
\end{split}
\end{equation}
$z_{P}=[1/c(k.p_{4})]\sqrt{(a_{1}.p_{4})^{2}+(a_{2}.p_{4})^{2}}$ is the argument of the ordinary Bessel functions expressed in the rest frame of the proton, and the phase $\phi_{0P}$ is defined by
\begin{equation}
\phi_{0P}=\arctan\bigg(\frac{a_{2}.p_{4}}{a_{1}.p_{4}}\bigg).
\end{equation}
The quantity $h'(Q_{3})$ is the derivative of the function $h(Q_{3})$ with respect to the energy $Q_{3}$ given by
\begin{equation}
h'(Q_{3})=\bigg[\frac{Q_{1}}{c^{2}}+M_{*}+\frac{(l+s)\omega}{c^{2}}\bigg]-\frac{Q_{3}}{c^{2}|\textbf{q}_{3}|}\bigg[|\textbf{q}_{1}|F(\theta_{i},\theta_{f},\varphi_{i},\varphi_{f})+\frac{(l+s)\omega}{c}\cos(\theta_{f})\bigg].
\end{equation}
\section{Results and discussion}\label{sec:res}
In this paper, we present a relativistic theoretical calculation, in the first Born approximation, to study the electron-proton scattering in the absence and the presence of a circularly polarized electromagnetic field. Before presenting the results and their physical interpretation, we would like to note that the various figures of the present work are plotted with the help of the programming language Mathematica. In this section, we will present all the numerical results obtained in both with and without laser; during that, we will follow the same arrangement that we adopted through the previous sections. We will begin with the results obtained in the absence of the laser field and then those obtained in its presence. All the DCSs are given in atomic units.
\subsection{In the absence of the laser field}
\begin{figure}[t]
\centering
\includegraphics[scale=0.7]{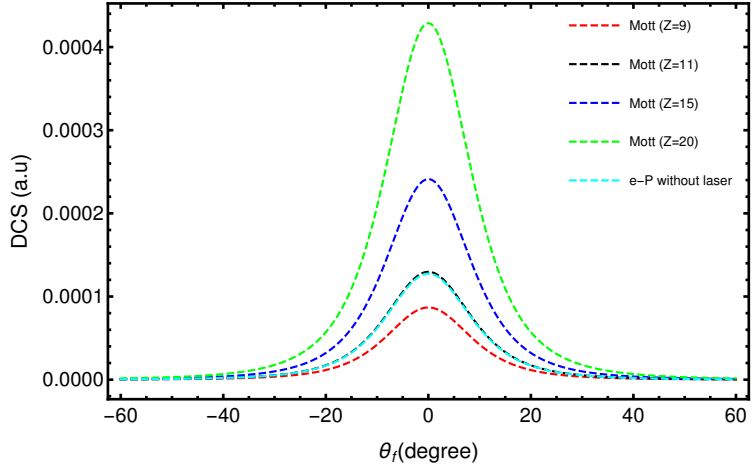}
\caption{The DCS of $e$-$P$ scattering without laser given in equation (\ref{dcswithout}) as a function of the scattering angle $\theta_{f}$ compared to that of Mott scattering for different atomic numbers $Z$. The electron kinetic energy is $T_{1}=0.511~\text{MeV}$. The geometry parameters are $\theta_{i}=\phi_{i}=15^{\circ}$, $\phi_{f}=105^{\circ}$.}\label{fig1}
\end{figure}

We will start our discussion in this section with a comparison between the two DCSs of electron-proton scattering and the well-known Mott scattering. The latter can be seen as a particular case of the former, in which the target is fixed and heavy so that it creates a Coulomb field around which all charged particles are affected. It has been found that the DCS expression of the electron-proton scattering becomes that of Mott scattering when the energy of the incoming electron is very small compared to the rest mass of the proton \cite{greiner}. At this limit, the proton does not recoil and can therefore be considered as a source of a constant external field. Conversely, when the electrons have high energy (the ultrarelativistic electrons) the recoil of the proton will inevitably change the expression of the DCS significantly. In the following, we will try to add some insights regarding the atomic number $Z$. In figure \ref{fig1}, we plot the DCS of the electron-proton scattering in the absence of the laser given in equation (\ref{dcswithout}) and compare it with that of Mott scattering at different atomic numbers. From this figure, we see that the DCS of Mott scattering increases with increasing atomic number $Z$. The shape of its curve is similar to that of a Gaussian function defined in the theory of statistic and probability as a normal distribution, with a peak at the final scattering angle $(\theta_{f}=0^{\circ})$. It is well-known that the DCS of Mott scattering, as expressed in theory, is quadratically dependent on the atomic number $Z$. When the atomic number $Z = 11$, we notice that the two DCSs of Mott and electron-proton scattering are equal. But, with the increase of the atomic number ($Z>11$ and therefore the target becomes very heavy), the Mott scattering becomes the most dominant and the valid method to study the process and vice-versa, which means that whenever the atomic number is lower than $11$ $(Z<11)$, we notice that the method used to analyse the Mott scattering is neglected and not valid in view of the recoil effect that appears at small values of atomic number (light atoms). Thus, the value $Z=11$ can be considered as a limit separating the two scatterings and determining the validity of one against the other.
\begin{figure}[t]
\centering
\includegraphics[scale=0.68]{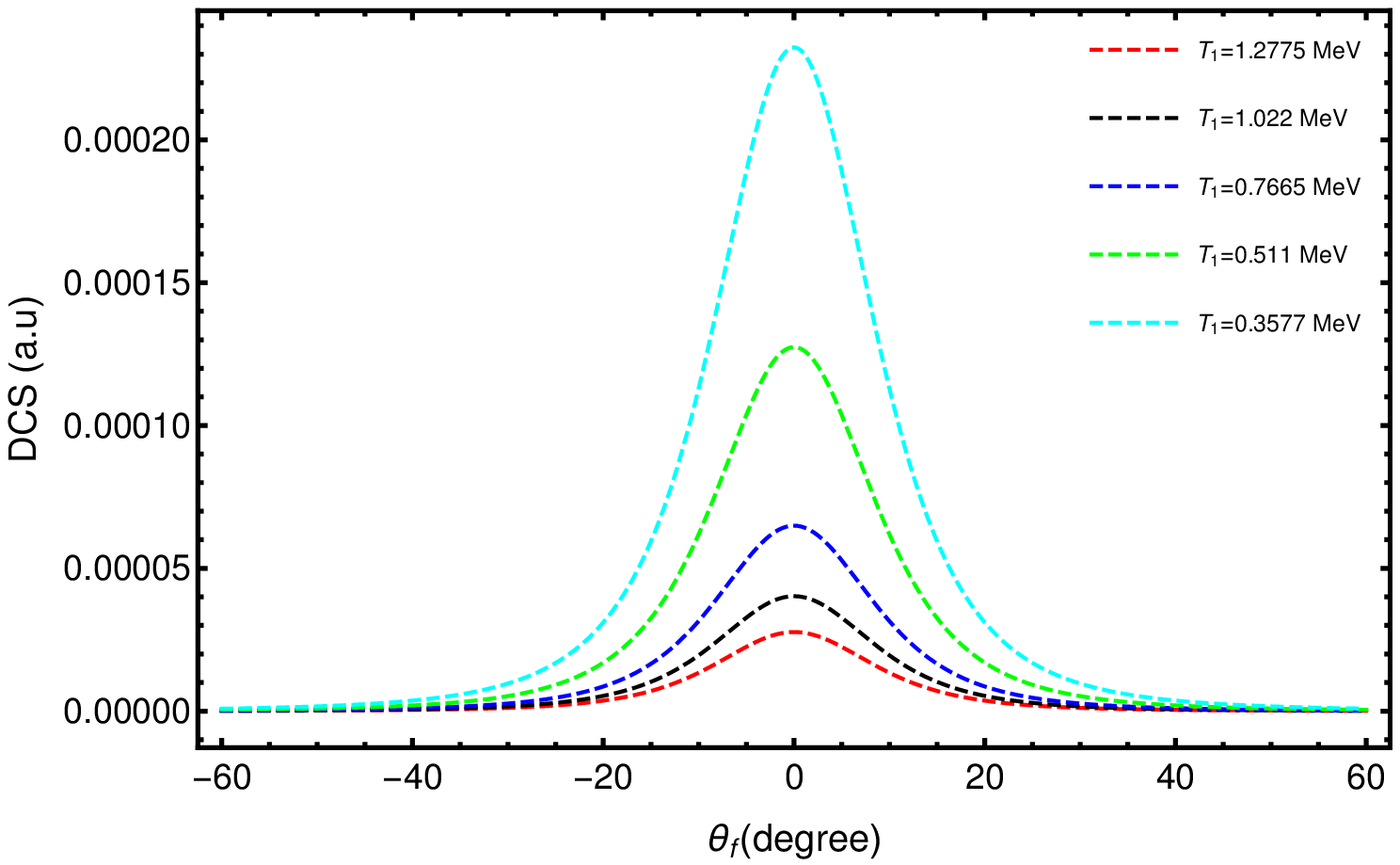}
\caption{The DCS of $e$-$P$ scattering without laser as a function of the scattering angle $\theta_{f}$ for different electron kinetic energies with the same geometry parameters  as in figure~\ref{fig1}.}\label{fig2}
\end{figure}
Figure \ref{fig2} shows the DCS changes of the electron-proton scattering without the presence of the laser in terms of scattering angle but at different kinetic energies. As expected, due to the unitarity of the S-matrix element, the DCS must be inversely proportional to the energy of the incoming electron, i.e. it decreases with increasing energy or vice-versa. The effects of relativity and spin have greatly contributed to these discrepancies that appear between both relativistic and nonrelativistic regimes. Just to get a complete and comprehensive overview of the DCS dependence on the incoming electron energy and its final scattering angle, we have inserted in figure \ref{fig3} a three-dimensional drawing (contour-plot) in which we represent the changes of the DCS in terms of two variables at the same time, the energy of the incoming electron and its final scattering angle $\theta_{f}$. From this figure, it can be seen that the peak of the curve  is located around the value $\theta_{f}=0$ and that the order of magnitude of the DCS decreases with increasing energy of the incoming electron.
\begin{figure}[hbtp]
\centering
\includegraphics[scale=0.7]{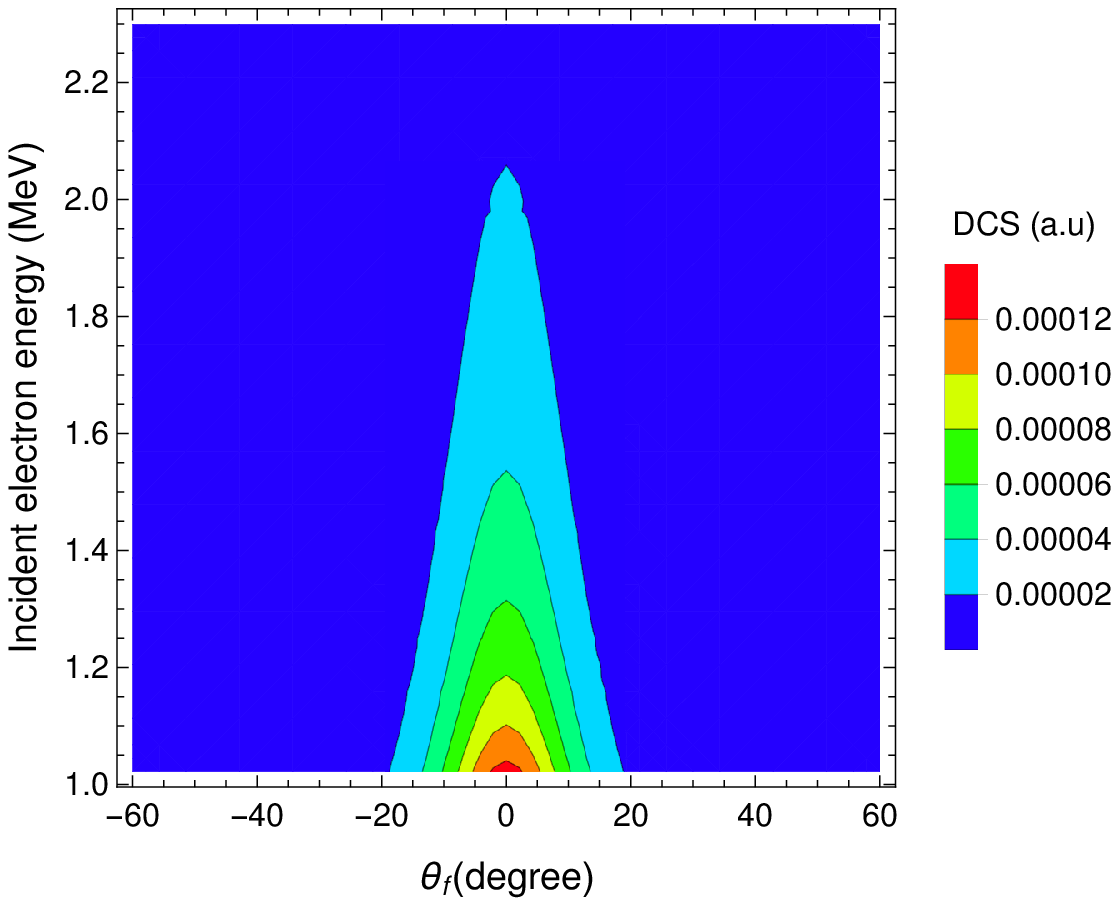}
\caption{The variations of the DCS of $e$-$P$ scattering without laser given in (\ref{dcswithout}) as a function of both incident electron energy $E_{1}$ and scattering angle $\theta_{f}$. The geometry parameters are the same as those in figure~\ref{fig1}.}\label{fig3}
\end{figure}
\begin{figure}[hbtp]
\centering
\includegraphics[scale=0.68]{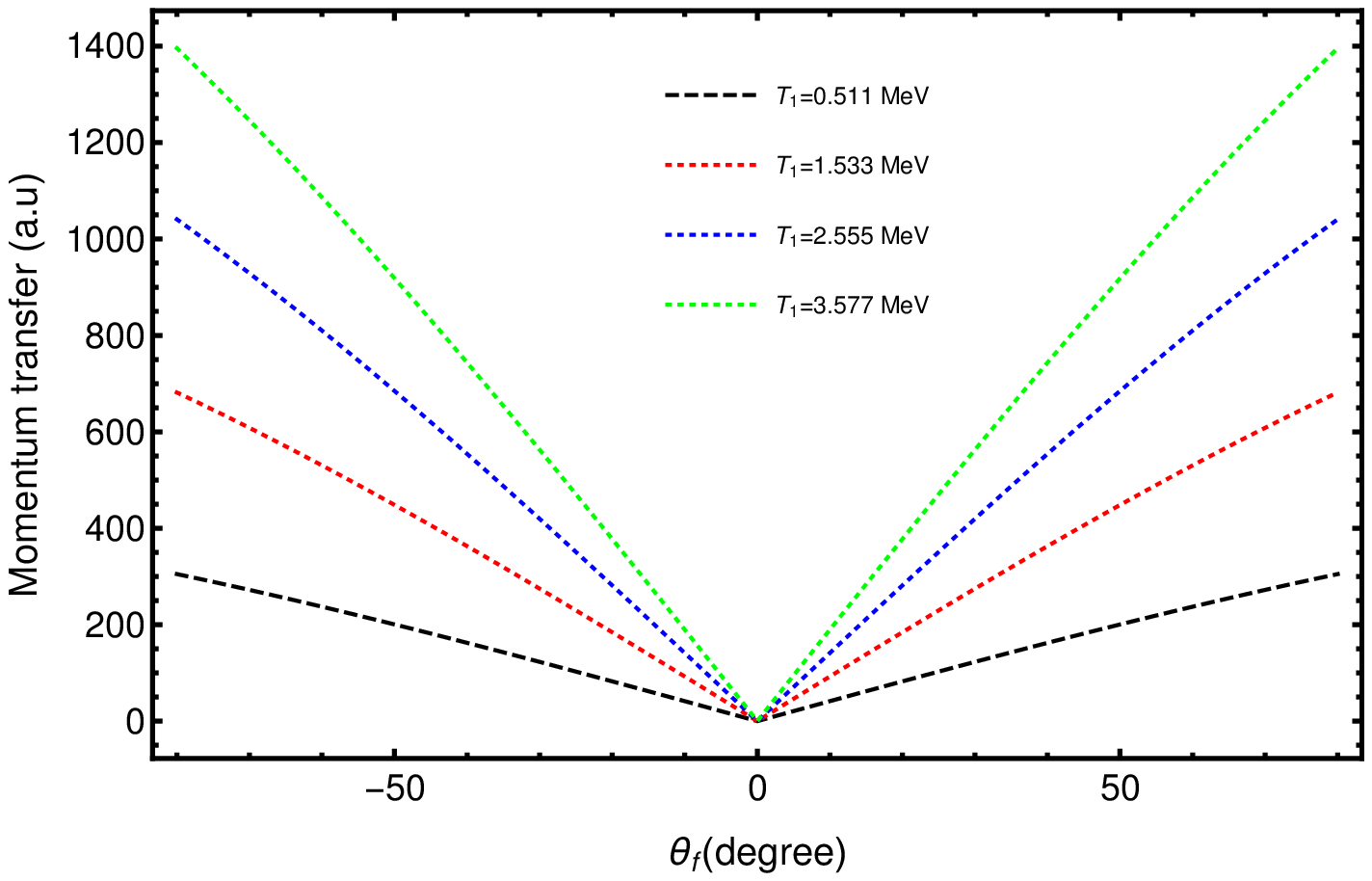}
\caption{The momentum transfer $|\textbf{q}|=|\textbf{p}_{3}-\textbf{p}_{1}|$ in the absence of the laser as a function of the scattering angle $\theta_{f}$ for different electron kinetic energies. The geometry parameters are $\theta_{i}=0^{\circ}$, $\phi_{i}=\phi_{f}=45^{\circ}$ and $-80\leq\theta_{f}\leq80$.}\label{fig4}
\end{figure}
We will end our discussion in this section by representing, in figure~\ref{fig4}, the changes in momentum transfer $|\textbf{q}|=|\textbf{p}_{3}-\textbf{p}_{1}|$ in the absence of a laser in terms of the scattering angle and for different kinetic energies. Momentum transfer is the basic physical quantity that characterizes all scattering theories and experiments, and it depends on the momentum of the projectile and the scattering angle. We note that the momentum transfer is a symmetric function with respect to the axis $\theta_{f}=0^{\circ}$ and its curve takes the lowest value at the same angle. It also increases almost linearly with the rise in kinetic energy and scattering angle. This behavior was clearly reflected in the DCS shown in the previous figures since the DCS, according to its expression in (\ref{dcswithout}), is inversely proportional to $q^{4}$. As we have seen previously, the DCS exhibits a strong peak at $\theta_{f}=0^{\circ}$ and falls off very rapidly at large momentum transfers.
\subsection{In the presence of the laser field}
\begin{figure}[t]
\centering
\includegraphics[scale=0.65]{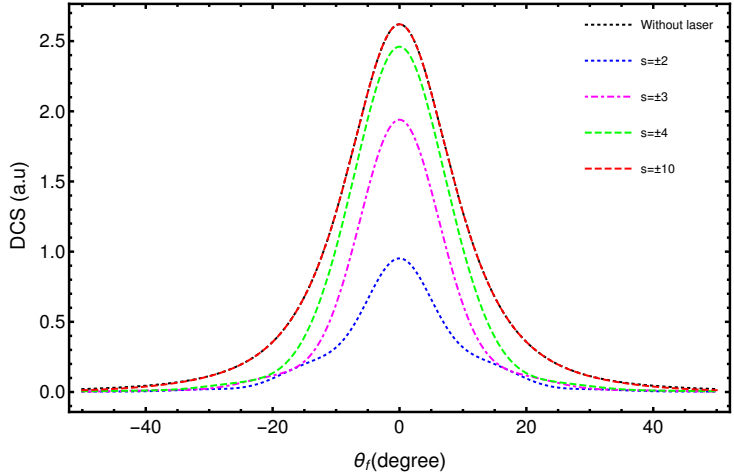}
\caption{The variations of the summed DCS of $e$-$P$ scattering with laser given in (\ref{dcswith}) as a function of the scattering angle $\theta_{f}$ for different numbers of photons exchanged. The laser field strength and frequency are $\mathcal{E}_{0}=10^{7}~\text{V/cm}$ and $\hbar\omega=1.17~\text{eV}$. The elecron kinetic energy is $T_{i}=2.7~\text{keV}$. The geometry parameters are $\theta_{i}=\phi_{i}=15^{\circ}$ and $\phi_{f}=105^{\circ}$. The notation $s=\pm N$ means that we have summed over $-N\leq s\leq+N$.}\label{fig5}
\end{figure}

\begin{figure}[hbtp]
\centering
\includegraphics[scale=0.6]{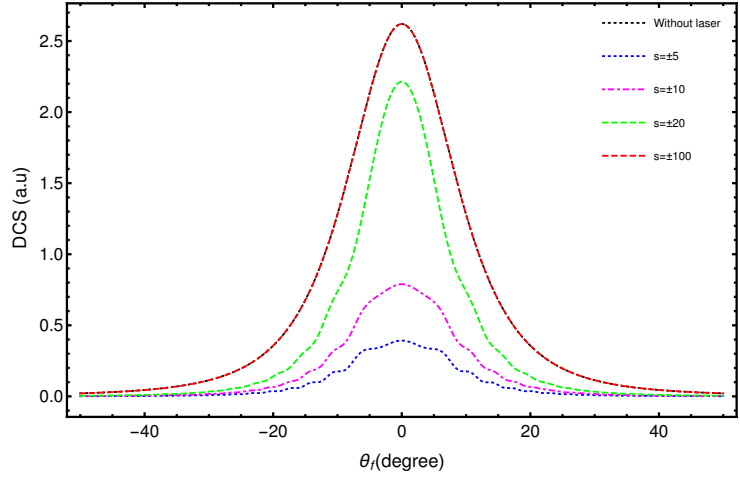}
\caption{The same as in figure \ref{fig5}, but for the laser field strength $\mathcal{E}_{0}=5.14225\times10^{7}~\text{V/cm}$. The notation $s=\pm N$ means that we have summed over $-N\leq s\leq+N$.}\label{fig6}
\end{figure}

\begin{table}[hbtp]
\centering
\caption{Values of $d\overline{\sigma}^{s}/d\Omega_{f}$ as a function of the number of photons exchanged $s$ at different field strengths. The laser frequency chosen is $\hbar\omega=1.17~\text{eV}$. The geometry parameters are $\theta_{i}=\phi_{i}=15^{\circ}$, $\theta_{f}=0^{\circ}$ and $\phi_{f}=105^{\circ}$.}
\begin{tabular}{|p{0.7cm}|p{3cm}|p{0.8cm}|p{4.5cm}|p{0.8cm}|p{3cm}|}
 \hline
~~$s$ & \shortstack{DCS (a.u)\\$\mathcal{E}_{0}=10^{7}~\text{V/cm}$\\$T_{1}=2.7~\text{keV}$}  & ~~$s$ & \shortstack{DCS (a.u)\\$\mathcal{E}_{0}=5.14225\times10^{7}~\text{V/cm}$\\$T_{1}=2.7~\text{keV}$} & ~~$s$ & \shortstack{DCS (a.u)\\$\mathcal{E}_{0}=10^{7}~\text{V/cm}$\\$T_{1}=0.511~\text{MeV}$} \\
 \hline
-10   & 0    &-100&  0 & -300& 0\\
-8   & 9.53968$\times10^{-5}$ &-50& $ 1.2521 \times10^{-27}$ & -200& 3.12079$\times10^{-7}$\\
-6   & 1.08317$\times10^{-2}$   &-20&   0.150387 & -100& 4.29275$\times10^{-7}$\\
-4   & 2.60332$\times10^{-1}$    &-10&  0.00463691 & -50& 1.25848$\times10^{-7}$\\
-2   & 2.35595$\times10^{-1}$   &-5&  0.0132268 & -20& 4.11216$\times10^{-7}$\\
~0   & 3.6098$\times10^{-1}$    &~0&  0.0224972 &~0& 9.43704$\times10^{-8}$\\
~2   & 2.35433$\times10^{-1}$   &~5&  0.0132039 & ~20& 4.11234$\times10^{-7}$\\
~4   & 2.59972$\times10^{-1}$   &~10&  0.00462091 & ~50& 1.25861$\times10^{-7}$\\
~6  & 1.08093$\times10^{-2}$    &~20&  0.149351  & ~100& 4.29368$\times10^{-7}$ \\
~8  & 9.53968$\times10^{-5}$    &~50& $ 1.23078 \times10^{-27}$ & ~200& 3.12214$\times10^{-7}$\\
~10  & 0   &~100& 0 & ~300& 0\\
 \hline
\end{tabular}
\label{tab1}
\end{table}
In this section, we will try to show what happens when we embed the electron-proton scattering in a circularly polarized electromagnetic field and what changes can occur in the DCS when the electromagnetic field is present. The latter is characterized by its parameters which are field strength and frequency. We will present here the results obtained in the case where we only dress the electron without the proton. The first thing we are accustomed to check in such processes occurring in the electromagnetic field is to make sure that the DCS in the presence of the laser field is exactly equal to the corresponding one in the absence of the laser when we take the zero field strength $(\mathcal{E}_{0}=0~\text{a.u})$ and without photon exchange $(s = 0)$. We give here two values of the DCS at $(\mathcal{E}_{0}=0~\text{a.u}~\text{and}~s=0)$, DCS$=2.61973~\text{a.u}$ at $T_{1}=2.7~\text{keV}$ and DCS$=1.27435\times10^{-4}~\text{a.u}$ at $T_{1}=0.511~\text{MeV}$, for the geometry parameters: $\theta_{i}=\phi_{i}=15^{\circ}$, $\theta_{f}=0^{\circ}$ and $\phi_{f}=105^{\circ}$. Table \ref{tab1} contains the values of the DCS $(d\overline{\sigma}^{s}/d\Omega_{f})$ as a function of the number of photons exchanged $s$ at different field strengths. We have chosen the laser frequency at $\hbar\omega=1.17~\text{eV}$. According to these values, it can be said that the number of exchanged photons enhances when the strength of the electromagnetic field $\mathcal{E}_{0}$ is increased. This is evident by the cutoff number which equals $\pm 10$ ($s=-10$ photons for the negative part and $s=+10$ photons for the positive part) in the case of $\mathcal{E}_{0}=10^{7}~\text{V/cm}=0.002~\text{a.u}$, while it is equal to $\pm100$ in the case of $\mathcal{E}_{0}=5.14225\times10^{7}~\text{V/cm}=0.01~\text{a.u}$. Regarding the order of magnitude of DCS, we can see that it decreases with increasing field strength. We can also conclude from the table that the DCS shows a symmetry of values with respect to $s=0$, indicating that the photon absorption processes $(s > 0)$ are exactly the same as the photon emission processes $(s < 0)$. When we raise only the kinetic energy $T_{1}$ of the incoming electron from $T_{1}=2.7~\text{keV}$ to $T_{1}=0.511~\text{MeV}$ and maintain the field strength at $\mathcal{E}_{0}=10^{7}~\text{V/cm}$, we notice that the multiphoton process becomes more important and the cutoff number rises to the value $s=\pm300$. This shows that due to the mobility property of the target, the photon transmission becomes more feasible and the DCS becomes smaller than the laser-free case. In figure \ref{fig5}, we have plotted the variations of the summed DCS given in (\ref{dcswith}) as a function of the scattering angle $\theta_{f}$ for different numbers of photons exchanged. The numbers of exchanged photons that we have summed over it are as follows: $s=\pm2, \pm3, \pm4, \pm10$, where the notation $s=\pm N$ means that we have summed over $-N\leq s\leq+N$. This figure shows that at $s=\pm10$, the DCS (with laser) is exactly close to the DCS (without laser). But for example, at $s=\pm2$, we have several orders of magnitude due to the difference between the two approaches and the case without laser is always the highest. Returning to the first case $s=\pm10$, the convergence achieved here proves the validation of the sum rule demonstrated by Kroll and Watson \cite{sumrule}. Based on this figure, we notice that the DCS with laser tends to approach that of the laser-free with increasing the number of exchanged photons until it is exactly equal to it at a specific number of photons called the cutoff number in which the sum rule is fulfilled. The cutoff number is defined as a determined number of photons from which the DCS $(d\overline{\sigma}^{s}/d\Omega_{f})$ falls abruptly ($d\overline{\sigma}^{s}/d\Omega_{f}=0~\text{a.u}$), which can be attributed to the well-known properties of the Bessel function, which diminishes strongly when its argument $z$ (equation~(\ref{argument})) is approximately equal to its order $s$. The cutoff number, according to the table \ref{tab1}, is $s=\pm10$ at $\mathcal{E}_{0}=10^{7}~\text{V/cm}$ and $\hbar\omega=1.17~\text{eV}$. In the non-relativistic regime, there is a rapid convergence towards the DCS without laser. The situation is difficult and becomes more complicated in the relativistic regime due to the lack of high-speed computers, and therefore we cannot sum over a very large number of exchanged photons. Hence, as we are limited in our computing capabilities, the convergence cannot be achieved numerically in this case. For instance, at $T_{1}=0.511~\text{MeV}$ it is necessary to sum over number of photons $-300\leq s\leq+300$ to recover the laser-free result. The sum rule implies that under particular conditions the DCS summed over all  exchanged photons, in the presence of a laser field, is equal to the field-free DCS for the same scattering process, taken at the same scattering angle and initial energy \cite{sumrule}.  The same applies to figure \ref{fig6}, where the sum rule is fulfilled at the photon number $\pm100$, which, according to table \ref{tab1}, is the cutoff number corresponding to the case where the electromagnetic field strength is $\mathcal{E}_{0}=5.14225\times10^{7}~\text{V/cm}$ and the frequency is $\hbar\omega=1.17~\text{eV}$. The sum rule can be, mathematically, represented by the following formula:
\begin{equation}
\sum_{s=-\text{cutoff}}^{+\text{cutoff}} \frac{d\overline{\sigma}^{s}}{d\Omega_{f}}=\bigg(\frac{d\overline{\sigma}}{d\Omega_{f}}\bigg)^{\text{laser-free}}.
\end{equation}
\begin{figure}[hbtp]
\centering
\includegraphics[scale=0.65]{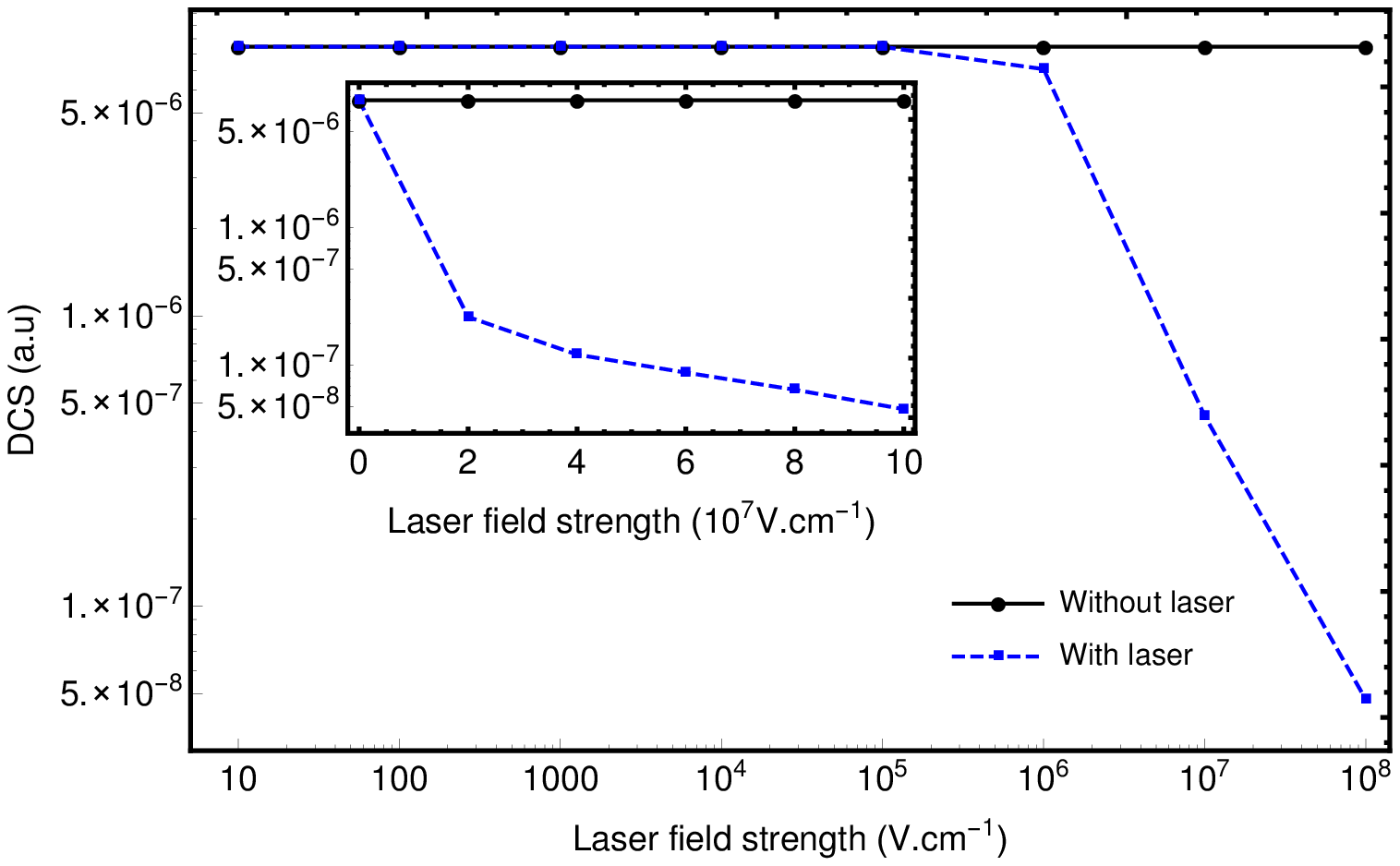}
\caption{The summed DCS of the $e$-$P$ scattering versus the laser field strength for the electron kinetic energy $T_{i}=0.0511~\text{MeV}$ and for an exchange of $\pm5$ photons at $\theta_{f}=90^{\circ}$ and $\theta_{i}=\phi_{i}=\phi_{f}=0^{\circ}$. The laser frequency is $\hbar \omega=1.17~\text{eV}$. The inset gives a clear zoom of the DCS variations for the field strengths between $10^{7}$ and $10^{8}~\text{V/cm}$.}\label{fig7}
\end{figure}
Figure \ref{fig7} shows the summed DCS dependence on the laser field strength. The inset shows more details of the same figure at field strengths between $10^{7}$ and $10^{8}~\text{V/cm}$. The field strength $\mathcal{E}_{0}$ is involved to determine the DCS behavior via the argument $z$ of Bessel functions in equation (\ref{spinorpart}). Although the Bessel functions in the partial DCS of each multiphoton process $(d\overline{\sigma}^{s}/d\Omega_{f})$ oscillate with the field strength, the summed DCS decreases gradually by orders of magnitude as the field strength increases from $10^{5}$ to $10^{8}~\text{V/cm}$. At weak filed strengths, i.e. between $10$ and $10^{5}~\text{V/cm}$, and with regard to the frequency $\hbar \omega=1.17~\text{eV}$, we notice that the laser has no effect on the DCS. However, the stronger the field strength and the more distortion of the electron state, the greater the reduction of the DCS. This means that the probability of scattering between the electron and proton is diminished in the presence of a circularly polarized laser field. This behavior has been verified on different geometries. The first simple physical explanation to which we can attribute this behavior is as follows: the electron, when embedded in a circularly polarized laser field, of course moves in perpendicular electric and magnetic fields, but at the same time rotates on an orbit whose radius corresponds to the value of the laser field strength. In contrast, the results obtained in the case of the linearly polarized laser field show that the laser makes a significant contribution to the enhancement of the DCS \cite{wang2019,liu2014}. In this case, the electron moves only in perpendicular electric and magnetic fields without any other movement, so that the probability of the electron being scattered with the proton will be greater with respect to the linearly polarized laser field than that corresponding to a circularly polarized laser field. Since this probability is directly related to the concept of the differential cross section in scattering theory, the results obtained in both cases are consistent with this interpretation. In figure \ref{fig8}, we seek to illustrate the effect of laser frequency on the DCS. To this end, we have plotted its changes in terms of laser field strength for two different frequencies, $\hbar\omega=0.117~\text{eV}$ and $\hbar\omega=1.17~\text{eV}$. As can be seen from figure \ref{fig8}, the low laser-frequency affects the DCS when the field strength exceeds the threshold of $10^{3}~\text{V/cm}$, while the effect of a high laser-frequency on the DCS does not appear until the field strength is above $10^{5}~\text{V/cm}$. Therefore, the first laser, which has a lower frequency, can affect the DCS more than the second one even at lower field strengths. Hence, the effect of the laser on the DCS becomes weak at high frequencies; and in order to make a high laser-frequency affecting the DCS, a very high field strengths are needed. For the sake of further illustration, like the 2-dimensional graphs, the contour graph in figure (\ref{fig9}) shows more information on the global variation and aspect of the DCS with respect to both laser field strength and frequency for the exchanged photons of about $\pm10$. In this type of 3D-graphics, it is the colors and contours that represent the changes in DCS on a two-dimensional plane. The DCS values are shown on the right of the graphic on the bar legend. According to the evolution of the laser frequencies, the effect of the laser field on the DCS decreases at high frequencies, while it becomes relevant as the laser field strength increases at each fixed frequency. This is perfectly consistent with all that we have discussed so far.
\begin{figure}[hbtp]
\centering
\includegraphics[scale=0.66]{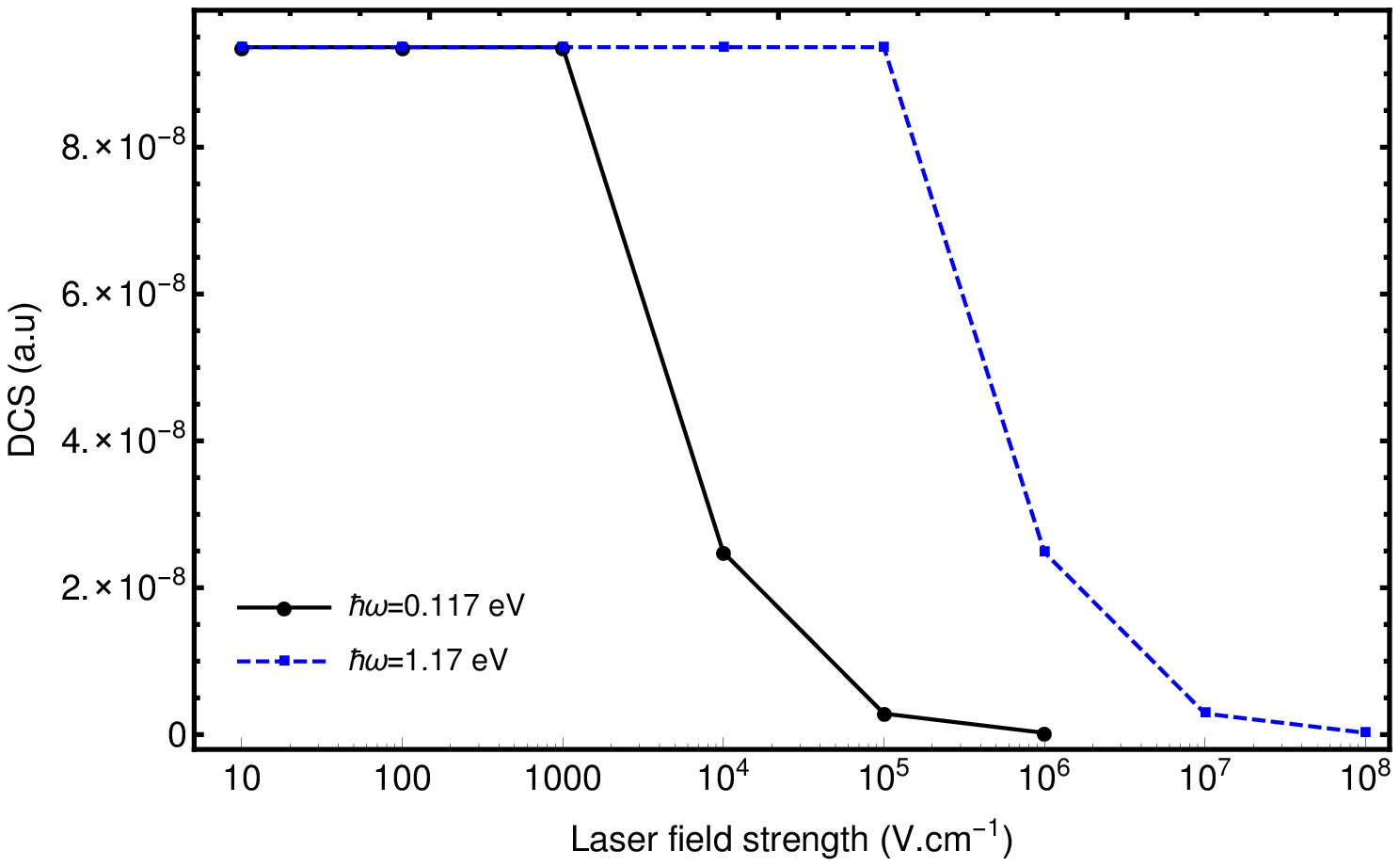}
\caption{The variations of the DCS of $e$-$P$ scattering with laser given in (\ref{dcswith}) as a function of laser field strength for two different frequencies $\hbar\omega=0.117~\text{eV}$ and $\hbar\omega=1.17~\text{eV}$. The numbers of photons exchanged is $s=\pm5$. The elecron kinetic energy is $T_{i}=0.511~\text{MeV}$. The geometry parameters are $\theta_{i}=\phi_{i}=\phi_{f}=0^{\circ}$ and $\theta_{f}=90^{\circ}$.}\label{fig8}
\end{figure}
\begin{figure}[hbtp]
\centering
\includegraphics[scale=0.66]{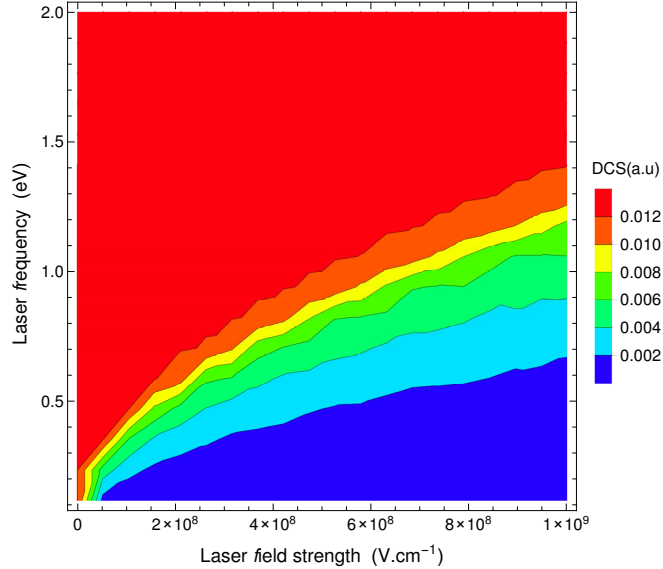}
\caption{The variations of the DCS of $e$-$P$ scattering with laser given in (\ref{dcswith}) as a function of both laser field strength and frequency for an exchange of $\pm10$ photons. The elecron kinetic energy is $T_{i}=2.7~\text{keV}$. The geometry parameters are $\theta_{i}=60^{\circ}$, $\theta_{f}=\phi_{i}=0^{\circ}$ and $\phi_{f}=90^{\circ}$. }\label{fig9}
\end{figure}
\subsection{Proton-dressing effect}
\begin{table}
\centering
\caption{Values of the two DCSs with and without the dressing effects of proton which are given, respectively, in equations (\ref{dressedprotondcs}) and (\ref{dcswith}) for different field strengths. The laser frequency is $\hbar\omega=1.17~\text{eV}$. The kinetic energy of the incoming electron is $T_{1}=2.7~\text{keV}$. The geometry parameters are $\theta_{i}=60^{\circ},\theta_{f}=1^{\circ},\phi_{i}=0^{\circ},\phi_{f}=90^{\circ}$. The two numbers of photons $s$ and $l$ over which we have summed are $-5\leq s\leq +5$ and $-5\leq l\leq +5$. The DCS of Mott scattering is also included for comparison.}
\begin{tabular}{|p{3cm}|p{3.7cm}|p{4.3cm}|p{4cm}|}
 \hline
Field strength $\mathcal{E}_{0}$ ~~~(V/cm)  & $\big(\frac{d\overline{\sigma}}{d\Omega_{f}}\big)^{\text{e$^-$-dressing}(\ref{dcswith})}$(a.u)  & $\big(\frac{d\overline{\sigma}}{d\Omega_{f}}\big)^{\text{(e$^-$,P)-dressing}(\ref{dressedprotondcs})}$(a.u)  & Mott DCS (a.u) with laser $(Z=1)$ \\
 \hline
$10^{7}$ & $0.00350813$  & $0.00350813$ &$0.0000294973 $ \\
$10^{8}$ & $0.000335896$ &$ 0.000335896$ & $2.8358\times10^{-6}$\\
$10^{9}$ &$ 0.0000440559$ & $0.0000440559$  & $3.80307\times10^{-7}$\\
$10^{10}$ &$ 1.34809\times10^{-9} $& $2.22418\times10^{-10}$ & $1.04815\times10^{-10}$\\
$10^{12}$ & $8.05724\times10^{-18}$ & $1.35211\times10^{-22}$ &$1.83393\times10^{-18}$ \\
$10^{13}$ & $2.034\times10^{-20}$ & $3.88512\times10^{-27}$ & $1.27317\times10^{-21}$\\
$10^{14}$ & $2.17776\times10^{-21}$ & $8.05327\times10^{-32}$ &$1.17667\times10^{-24}$ \\
  \hline
\end{tabular}
\label{tab2}
\end{table}
This section is reserved for the presentation and discussion of the results obtained in the case of the dressed proton. When we consider the interaction of both electron and proton with the electromagnetic field in the initial and final states, the theoretical calculation becomes difficult and somewhat cumbersome. The extraction of numerical and graphical results related to this situation requires high-speed and high-resolution computing facilities. Summing over two numbers of exchanged photons, one for the electron $(s)$ and the other for the proton $(l)$, takes time and effort twice. Therefore, we will limit ourselves here to include, in table \ref{tab2}, a few relevant values for the two DCSs in case we dress only electron or  electron and proton together. Looking at table \ref{tab2}, it is easy to see that the DCSs of the two cases are equal at the first three field strengths $(10^{7},~10^{8}$ and $10^{9}~\text{V/cm})$, which means that the proton dressing has no effect yet. Once we are above field strength $10^{9}~\text{V/cm}$, we notice that the two DCSs begin to differ significantly. These values justified the fact that the interaction of the proton with the laser field at field strengths between $0$ and $10^{9}~\text{V/cm}$ was neglected in this work (section \ref{secwith}) as well as in previous works related to this subject \cite{wang2019,liu2014}. Within this interval of strengths, no matter whether the proton dressing is considered or not, it is the same and no change occurs. But when higher field strengths are reached outside this interval $(\mathcal{E}_{0}\geq 10^{10}~\text{V/cm})$, the proton dressing must be taken into account. This can reasonably be explained by the heavier mass of the proton compared to that of the electron. From table \ref{tab2}, it can be seen that the inclusion of the proton dressing has contributed to a significant reduction in DCS more than before ; and the nature of the electromagnetic field polarization plays an important role in this behavior. Comparing these two DCSs with that of Mott scattering at $Z=1$, we can see that, due to the recoil effect, there is a deviation between them, and that of Mott is always smaller.
\section{Conclusion}\label{sec:conclusion}
In this work, we have studied and revealed the effects of the proton dressing in the relativistic electron-proton scattering in the presence of a circularly polarized monochromatic laser field. We started with checking the results already known for electron-proton scattering in the absence of the laser field. After that, we moved to the case in which we apply the laser field, either to only the electron or to the electron and proton together. Our theoretical results show that the DCS of electron-proton scattering is greatly minimized by the presence of the laser field. In the case of the proton dressing, we have found that it has no effect at all as long as the field strength is below $10^{10}~\text{V/cm}$ and can therefore be neglected. Despite all of this, we emphasize that the approach presented here may not provide a realistic description of electron-proton scattering at very high energies. In this case, the models considering proton as a structureless particle are no longer accurate, and it has therefore become necessary to introduce the form factors responsible for the substructure of the proton (deep inelastic scattering). In any case, it seems that this issue is of particular interest and deserves a future investigation.


\begin{thebibliography}{99}
\bibitem{highlaser} Bahk S W, Rousseau P, Planchon T A, Chvykov V, Kalintchenko G, Maksimchuk A, Mourou G A and Yanovsky V 2004 Generation and characterization of the highest laser intensities ($10^{22}~\text{W/cm}^{2}$) \emph{Opt. Lett.} \textbf{29} 2837
\bibitem{faisal} Faisal F H M 1987 \textit{Theory of Multiphoton Processes} (New York:
Plenum)
\bibitem{mittleman} Mittleman M H 1993 \textit{Introduction to the Theory of Laser-Atom Interactions} (New York: Plenum)
\bibitem{fedorov} Fedorov M V 1997 \textit{Atomic and Free Electrons in a Strong Light Field} (Singapore: World Scientific)
\bibitem{review2019} V\'elez F C, Kaminski J Z and Krajewska K 2019 Electron scattering processes in non-monochromatic and relativistically intense laser fields \emph{Atoms} \textbf{7} 34
\bibitem{review2009} Ehlotzky F, Krajewska K and Kaminski J Z 2009 Fundamental processes of quantum electrodynamics in laser fields of relativistic power \emph{Rep. Prog. Phys.} \textbf{72} 046401
\bibitem{review1990} Francken P and Joachain C J 1990 Theoretical study of electron-atom collisions in intense laser fields \emph{J. Opt. Soc. Am. B} \textbf{7} 554
\bibitem{review1998} Ehlotzky F, Jaron A and Kaminski J Z 1998 Electron-atom collisions in a laser field \emph{Phys. Rep.} \textbf{297} 63
\bibitem{nospin1} Kaminski J Z and Ehlotzky F 1999 Asymmetries and dark angular windows in relativistic free-free transitions in a powerful laser field \emph{Phys. Rev. A} \textbf{59} 2105
\bibitem{nospin2} Panek P, Kaminski J Z and Ehlotzky F 1999 Angular and polarization effects in relativistic potential scattering of electrons in a powerful laser field \emph{Can. J. Phys.} \textbf{77} 591
\bibitem{nospin3} Ehlotzky F 1988 Sum rules in relativistic potential scattering in a strong laser field \emph{Opt. Commun.} \textbf{66} 265
\bibitem{szymanowski1} Szymanowski C and Maquet A 1998 Relativistic signatures in
laser-assisted scattering at high field intensities \emph{Opt. Express} \textbf{2} 262
\bibitem{szymanowski2} Szymanowski C, V\'eniard V, Ta\"{i}eb R, Maquet A and Keitel C H 1997 Mott scattering in strong laser fields \emph{Phys. Rev. A} \textbf{56} 3846
\bibitem{li2003} Li S M, Berakdar J, Chen J and  Zhou Z F 2003 Mott scattering in the presence of a linearly polarized laser field \emph{Phys. Rev. A} \textbf{67} 063409
\bibitem{li2004} Li S M, Berakdar J, Chen J and  Zhou Z F 2004 Laser-assisted Mott scattering in the Coulomb approximation \emph{J. Phys. B: At. Mol. Opt. Phys.} \textbf{37} 653
\bibitem{attaourti2004} Attaourti Y,  Manaut B and Taj S 2004 Mott scattering in an elliptically polarized laser field \emph{Phys. Rev. A} \textbf{70} 023404
\bibitem{manaut2005} Manaut B, Taj S and Attaourti Y 2005 Mott scattering of polarized electrons in a strong laser field \emph{Phys. Rev. A} \textbf{71} 043401
\bibitem{manaut2009} Manaut B, Attaourti Y, Taj S and Elhandi S 2009 Mott scattering of polarized electrons in a circularly polarized laser field \emph{Phys. Scr.} \textbf{80} 025304
\bibitem{panek2004} Panek P, Kaminski J Z and Ehlotzky F 2004 Analysis of resonances in M\o ller scattering in a laser field of relativistic radiation power \emph{Phys. Rev. A} \textbf{69} 013404
\bibitem{bhabha1} Denisenko O I and Roshchupkin S P 1999 Resonant scattering of an electron by a positron in the field of a light wave \emph{Laser Phys.} \textbf{9} 1108
\bibitem{bhabha2}  Roshchupkin S P, Voroshilo A I and Padusenko E A 2010 One photon annihilation of an electron-positron pair in the field of pulsed circularly polarized light wave \emph{Laser Physics} \textbf{20} 1679
\bibitem{du2018} Du W Y, Zhang P F and Wang B H 2018 New phenomena in laser-assisted scattering of an electron by a muon \emph{Front. Phys.} \textbf{13} 133401
\bibitem{muon1} Padusenko E A, Roshchupkin S P and Voroshilo A I 2009 Nonresonant scattering of relativistic electron by relativistic muon in the pulsed light field \emph{Laser Phys. Lett.} \textbf{6} 242
\bibitem{muon2} Nedoreshta V N, Voroshilo A I and Roshchupkin S P 2008 Resonant scattering of an electron by a muon in the field of light wave \emph{Eur. Phys. J. D} \textbf{48} 451
\bibitem{muon3} Nedoreshta V N, Voroshilo A I and Roshchupkin S P 2007 Nonresonant scattering of an electron by a muon in the field of plane electromagnetic wave \emph{Laser Phys. Lett.} \textbf{4} 872
\bibitem{jakha} Jakha M, Mouslih S, Taj S and Manaut B 2021 Laser effect on the final products of $Z$-boson decay \emph{Laser Phys. Lett.} \textbf{18} 016002
\bibitem{mouslih} Mouslih S, Jakha M, Taj S, Manaut B and Siher E 2020 Laser-assisted pion decay \emph{Phys. Rev. D} \textbf{102} 073006
\bibitem{dicus1} Dicus D A, Farzinnia A, Repko W W and Tinsley T M 2009 Muon decay in a laser field \emph{Phys. Rev. D} \textbf{79} 013004
\bibitem{dicus2} Farzinnia A, Dicus D A, Repko W W and Tinsley T M 2009 Muon decay in a linearly polarized laser field \emph{Phys. Rev. D} \textbf{80} 073004
\bibitem{liu2007} Liu A H, Li S M and Berakdar J 2007 Laser-assisted muon decay \emph{Phys. Rev. Lett.} \textbf{98} 251803
\bibitem{neutrino} Bai L, Zheng M Y and Wang B H 2012 Multiphoton processes in laser-assisted scattering of a muon neutrino by an electron \emph{Phys. Rev. A} \textbf{85} 013402
\bibitem{brems1} L\"{o}tstedt E, Jentschura U D and Keitel C H 2007 Evaluation of laser-assisted bremsstrahlung with Dirac-Volkov propagators \emph{Phys. Rev. Lett.}
\textbf{98} 043002
\bibitem{brems2} Schnez S, L\"{o}tstedt E, Jentschura U D and Keitel C H 2007 Laser-assisted bremsstrahlung for circular and linear polarization \emph{Phys. Rev. A} \textbf{75} 053412
\bibitem{pairprod1} M\"{u}ller C, Hatsagortsyan K Z and Keitel C H 2008 Muon pair creation from positronium in a linearly polarized laser field \emph{Phys. Rev. D} \textbf{78} 033408
\bibitem{pairprod2} M\"{u}ller C, Hatsagortsyan K Z and Keitel C H 2006 Muon pair creation from positronium in a circularly polarized laser field \emph{Phys. Rev. D} \textbf{74} 074017
\bibitem{rosenbluth} Rosenbluth M N 1950 High energy elastic scattering of electrons on protons \emph{Phys. Rev.} \textbf{79} 615
\bibitem{corrections1} Maximon L C and Parke W C 2000 Radiative corrections to elastic electron-proton scattering for polarized electrons \emph{Phys. Rev. C} \textbf{61} 045502
\bibitem{corrections2} Krass A S 1962 Radiative corrections to electron-proton scattering \emph{Phys. Rev.} \textbf{125} 2172
\bibitem{corrections3} Tsai Y S 1961 Radiative corrections to electron-proton scattering \emph{Phys. Rev.} \textbf{122} 1898
\bibitem{liu2014} Liu A H and Li S M 2014 Relativistic electron scattering from a freely movable proton in a strong laser field \emph{Phys. Rev. A} \textbf{90} 055403
\bibitem{wang2019} Wang N, Jiao L and Liu A 2019 Relativistic electron scattering from freely movable proton/$\mu^{+}$ in the presence of strong laser field \emph{Chin. Phys. B} \textbf{28} 093402
\bibitem{ion} Collins L A and Csanak G 1991 Multiphoton resonances in $e+\text{H}^{+}$ scattering in a linearly polarized radiation field \emph{Phys. Rev. A} \textbf{44} R5343
\bibitem{atom} Franz A, Klar H, Broad J T and Briggs J S 1990 Electron-proton scattering in a laser field \emph{J. Opt. Soc. Am. B} \textbf{7} 545
\bibitem{greiner} Greiner W and M\"{u}ller B 2000 \textit{Gauge Theory of Weak Interactions} (Berlin: Springer)
\bibitem{volkov} Volkov D M 1935 On a class of solutions of the Dirac equation \emph{Z. Phys.} \textbf{94} 250
\bibitem{landau} Berestetskii V B, Lifshitz E M and Pitaevskii L P 1982 \emph{Quantum Electrodynamics} (Oxford U.K.: Butterworth-Heinemann)
\bibitem{feyncalc1} Mertig R, B\"{o}hm M and Denner A 1991 Feyn Calc - Computer-algebraic calculation of Feynman amplitudes \emph{Comput. Phys. Commun.} \textbf{64} 345
\bibitem{feyncalc2} Shtabovenko V, Mertig R and Orellana F 2016 New developments in FeynCalc 9.0  \emph{Comput. Phys. Commun.} \textbf{207} 432
\bibitem{feyncalc3} Shtabovenko V, Mertig R and Orellana F 2020 FeynCalc 9.3: New features and improvements \emph{Comput. Phys. Commun.} \textbf{256} 107478
\bibitem{sumrule} Kroll N M and Watson K M 1973 Charged-particle scattering in the presence of a strong electromagnetic wave \emph{Phys. Rev. A} \textbf{8} 804
\end{thebibliography}
\end{document}